\documentclass[a4paper,11pt]{article}
\pdfoutput=1 

\usepackage{jheppub} 

\usepackage{epstopdf} 
\usepackage[T1]{fontenc} 
\usepackage[dvipsnames]{xcolor}

\usepackage{graphics}
\usepackage[utf8]{inputenc}
\usepackage{xspace}
\usepackage{amsmath}
\usepackage{amssymb}
\usepackage{url}
\usepackage{rotating}
\usepackage{enumerate}
\usepackage{graphicx}
\usepackage{pdflscape, afterpage, capt-of} 
\usepackage{slashed} 
\usepackage[normalem]{ulem}
\usepackage{nicefrac}
\usepackage[export]{adjustbox} 
\usepackage{makecell} 
\usepackage{subcaption} 
\usepackage{bbold}
\usepackage{stackengine} 

\usepackage{soul} 

\numberwithin{equation}{subsection}

\newcommand{\mga}{\textsc{MadGraph5\_aMC@NLO}}

\usepackage{def} 

\newcommand{\private}[1]{} 
\renewcommand{\commentMS}[1]{\private{\textcolor{blue}{\sl MS: #1}}}
\renewcommand{\commentZW}[1]{\private{\textcolor{PineGreen}{\sl ZW: #1}}}

\renewcommand{\ZW}[1]{\private{\commentZW{#1}}}

\makeatletter
\def\hlinewd#1{
\noalign{\ifnum0=`}\fi\hrule \@height #1 \futurelet
\reserved@a\@xhline}
\makeatother

\title{\boldmath Simplifying QCD event generation with chirality flow, reference vectors and spin directions}

\author[a]{Emil Boman,}
\author[a]{Andrew Lifson,}
\author[a]{Malin Sjodahl,}
\author[b]{Adam Warnerbring}
\author[c,d]{and Zenny Wettersten}

\affiliation[a]{Department of Physics, 
  Lund University, Box 118, 221 00 Lund, Sweden}
\affiliation[b]{Center for Particle Physics Siegen - Experimental Particle Physics, \\University of Siegen, Walter-Flex-Str. 3, 57072 Siegen, Germany}
\affiliation[c]{CERN, 1211 Geneva 23, Switzerland}
\affiliation[d]{Institute of High Energy Physics, Austrian Academy of Sciences, \\Nikolsdorfer Gasse 18, 1050 Vienna, Austria }

\abstract{

  The chirality-flow formalism, combined with good choices
  of gauge reference vectors, 
  simplifies tree-level calculations to the extent
  that it is often possible to write down amplitudes corresponding to Feynman
  diagrams immediately. It has also proven to give
  a very sizable speedup
  in a proof of concept implementation of massless tree-level QED in \mga.
  In the present paper we extend this analysis to QCD, including massive
  quarks.
  We define helicity-dependent versions of the gluon vertices,
  derive constraints on the spinor structure of propagating
  gluons,
  and explore the Schouten identity to simplify the four-gluon vertex
  further.
  For massive quarks, the chirality-flow formalism sheds light on
  how to 
  exploit the freedom to measure spin along any direction to shorten
  the calculations.
  Overall, this results in a clear speedup for treating the Lorentz
  structure at high multiplicities.

  }

\begin{document} 
\preprint{MCnet-23-20}
\maketitle
\flushbottom

\section{Introduction}
\label{sec:introduction}

Tools for simulating events at the LHC and other collider
experiments tend to employ Feynman diagram methods
to compute helicity amplitudes, i.e., amplitudes with
assigned helicities \cite{Gleisberg:2008fv,Alwall:2014hca,Sjostrand:2014zea,Bellm:2015jjp,Sherpa:2019gpd}. 
Such calculations, especially for massless particles,
benefit from utilizing the spinor-helicity formalism
\cite{Farrar:1983wk,Berends:1987cv,Berends:1987me,Berends:1988yn,Berends:1988zn,Berends:1989hf,Murayama:1992gi,Dittmaier:1993bj,Dittmaier:1998nn,Weinzierl:2005dd,Gleisberg:2008fv}.
For treating the color structure, orthogonal bases, based on  
representation theory could be used \cite{Keppeler:2012ih,Sjodahl:2015qoa,Du:2015apa,Alcock-Zeilinger:2022hrk}, but so
far, in practice typically trace or color-flow bases are employed
\cite{tHooft:1973alw, Kanaki:2000ms, Maltoni:2002mq,Weinzierl:2005dd,Kilian:2012pz,Sjodahl:2012nk,Reuschle:2013qna,Sjodahl:2014opa}.

In a few recent papers \cite{Lifson:2020pai,Alnefjord:2020xqr,Lifson:2023wow}
the Lorentz structure analogue of the
color-flow picture was developed, by also ``flowing''
the spin and momentum structure of amplitudes.
Similar to the color-flow Feynman rules, a complete set
of chirality-flow Feynman rules were developed for the
standard model in \cite{Alnefjord:2020xqr,Lifson:2023wow}.
These chirality-flow rules, 
where the Fierz identity has been built in (like the color Fierz
identity for color flow), is easily seen to
significantly simplify tree-level calculations by hand
\cite{Lifson:2020pai,Alnefjord:2020xqr}. Recently it has also proven
to give a sizable speedgain for a numerical tree-level
massless QED implementation within \mga\;\cite{Alwall:2014hca}, where
the computation time is reduced by more than a factor 10 for
$e^+e^-$ going to seven photons \cite{Lifson:2022ijv}.

This improvement is in part attributed to the simpler Lorentz
structure for the (massless) QED vertex, and in part due
the choice of gauge reference vector for external
photons. While polarization vectors typically are taken
to be orthogonal to the momentum $p$ and a four-vector with
the sign of the spatial part swapped, one may
perfectly well describe polarization vectors as being
orthogonal to some other (lightlike) reference momentum $r$, aside from $p$
\cite{Xu:1986xb, Mangano:1987xk}.

In this paper we address QCD 
and explore the simplifications and speedgains
achieved by clever choices of reference and spin vectors, combined 
with simplified versions of the chiral three- and four-gluon
vertices. In particular, this makes it possible to retain
information about the spinor structure of internal gluons.
In this context we also exploit the Schouten identity.

Although most strongly interacting particles effectively are
massless, the top quark requires a mass.
We therefore address the complications brought about by
adding fermion masses; while massless fermions
naturally lend themselves to a description in
terms of single Weyl spinors, massive fermions necessarily
come with both a left- and a right-chiral component,
which unavoidably complicates the calculations.
However, also in this case, a clever choice of the
reference vector --- which now has acquired a physical
interpretation, being related to the direction in
which spin is measured --- can be explored to reduce the
computational complexity. Effectively, we thereby take
advantage of the freedom to measure spin in any
direction in order to simplify calculations.

To set the stage of our implementation, we 
introduce the chirality-flow formalism in the following
section. After that, we describe the context within
which the chirality-flow version of QCD is implemented
in \secref{sec:fhelas}, where we outline the HELAS  algorithm \cite{Murayama:1992gi}
and its implementation within \mga.
The algebraic simplifications from gauge dependent Feynman rules
are explored in \secref{sec:QCD}, and
the resulting speedgains in \secref{sec:results}.
Finally, we conclude and discuss future perspectives
in \secref{sec:conclusion}.

\section{Chirality flow}
\label{sec:chirality flow}

By noting that the Lie algebra $\mathfrak{so}(1,3)$ of the restricted
Lorentz group SO(1,3) can be decomposed into two 
copies of the SL(2,$\mathbb{C}$) Lie algebra $\mathfrak{sl}(2,\mathbb{C})$,
the spinor-helicity \citep{DeCausmaecker:1981jtq,Berends:1981rb,
Berends:1981uq,DeCausmaecker:1981wzb,Berends:1983ez,Kleiss:1984dp,
Berends:1984gf,Gunion:1985bp,Gunion:1985vca,Kleiss:1986qc,Hagiwara:1985yu,Kleiss:1985yh,
Kleiss:1986ct,Xu:1986xb,Gastmans:1987qz,Schwinn:2005pi} and
Weyl-van-der-Waerden \citep{Farrar:1983wk,Berends:1987me,Berends:1987cv,
Berends:1988yn,Berends:1988zn,Berends:1989hf,Dittmaier:1993bj,
Dittmaier:1998nn,Weinzierl:2005dd} formalisms split Minkowski four-space
into two two-component spinor spaces. In chirality flow,
we further incorporate the Fierz identity for the Pauli matrices,
\begin{align}
\sigma^{\mu, \dot{\alpha} \beta} \bar{\sigma}_{\mu, \alpha \dot{\beta}} = 2
\delta^{\dot{ \alpha }}_{\ \dot{\beta}} \delta^{\ \beta}_{\alpha},
\end{align}
directly into our Feynman rules, analogous to how gluons are split into
color and anti-color lines in QCD \cite{tHooft:1973alw, Kanaki:2000ms, Maltoni:2002mq}. 
This gives rise to a flow-description of chiral representations of Feynman
rules and diagrams,
dubbed
the chirality-flow formalism
\citep{Lifson:2020pai,Alnefjord:2020xqr,Lifson:2023wow}.
Below we go through the essence of chirality-flow Feynman rules.
Our conventions
are extensively detailed in ref. \citep{Lifson:2020pai}.

\subsection{A graphical representation of fermions and gauge bosons}
\label{sec:spinors}

The fundamental building blocks for calculations in the spinor helicity formalism
in general are the 
left- and right-chiral Weyl spinors, which we graphically represent as
\begin{alignat}{2}
 \lanSp{i} &=
 \raisebox{-0.3\height}{\includegraphics[scale=0.4]{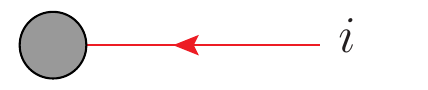}},
 \qquad  
 \sqlSp{i} &&=
 \raisebox{-0.3\height}{\includegraphics[scale=0.4]{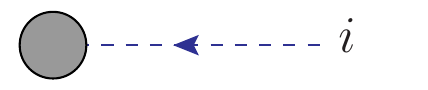}},
 \nonumber \\
 \ranSp{j} &=
 \raisebox{-0.3\height}{\includegraphics[scale=0.4]{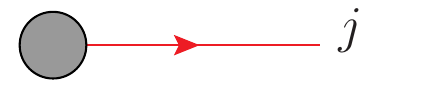}}, 
 \qquad 
 \sqrSp{j} &&=
 \raisebox{-0.3\height}{\includegraphics[scale=0.4]{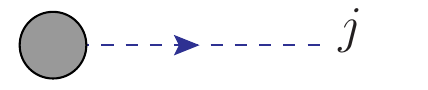}},
\label{eq:all spinors}
\end{alignat}
where $\lanSp{i}=\lanSp{p_i}$ etc. We define the direction of chirality
flow such that following the arrow direction is equivalent to
reading from left to right in an equation. Thus the Lorentz invariant spinor
inner products are given by
\begin{equation}
  \lan i j \ran 
  \defequal  
  \epsilon^{\al \be}\la_{i,\be}\la_{j,\al}
   = 
   {\includegraphics[scale=0.4]{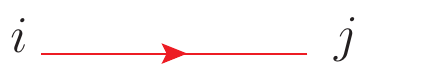}} 
  ~, 
   \;\;\;
  \sql  i j  \sqr 
  \defequal
  \epsilon_{\da \db}\tla_i^{\db}\tla_j^{\da}
  = \raisebox{-0.2\height}{\includegraphics[scale=0.4]{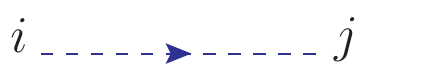}} ~,
    \label{eq:inner products}
\end{equation}
with (antisymmetric) $\lan ij \ran \sim \sql ij \sqr \sim
\sqrt{2p_i\cdot p_j}$, up to a phase factor.

While there is a simple relationship between massless fermions 
and Weyl spinors, e.g.
\begin{align}
u^+(p) \; = \; \begin{pmatrix}
0
\\
\raisebox{-5.5pt}{\includegraphics[scale=0.375]{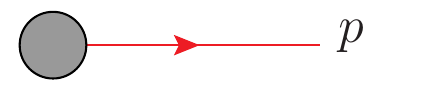}}
\end{pmatrix}
\rightarrow \raisebox{-5.5pt}{\includegraphics[scale=0.375]
{./Jaxodraw/MS/ExtSpinorAntiSolid_pCol}}, \quad p^2 = 0,
\end{align}
we also need to describe vector bosons and
massive fermions. In order to treat massive momenta $p$, we decompose
them in terms of two lightlike momenta $p^{\flat}$ and $q$ by
\begin{equation}
p^{\mu} = p^{\flat,\mu}+\alpha q^\mu~, \qquad (p^\flat)^2=q^2=0, \qquad p^2=m^2~,
\qquad \alpha = \frac{m^2}{2p^\flat\cdot q} = \frac{m^2}{2p\cdot q}~.
\label{eq:massive_momentum_decomposition}
\end{equation}
Note that \eqref{eq:massive_momentum_decomposition} does not determine the
decomposition $p = \pflat + \alpha q$ uniquely. This is related to the freedom to measure spin along
any axis \citep{Kleiss:1985yh,Dittmaier:1998nn},
\begin{align}
  \label{eq:spin_vec}
  s^{\mu} = \frac{1}{m} \left( \pflatmu - \alpha q^{\mu} \right) =
  \frac{1}{m} \left( p^{\mu} - 2\alpha q^{\mu} \right).
\end{align}
Measuring spin along
the direction of motion (corresponding to
$q = \frac{1}{2}(p_0- |\vec{p}|)(1,-\frac{\vec{p}}{|\vec{p}|})$, $\alpha=1$,
$s=\frac{1}{m}(|\vec{p}|, p^0 \frac{\vec{p}}{|\vec{p}|})$, cf. \cite{Alnefjord:2020xqr}) reproduces the helicity basis, \commentMS{New:check} 
but we
will instead keep the spin direction arbitrary here, such that we can
use it to reduce the complexity of the computations,
a topic we discuss further in \secref{sec:spin and gauge Feyn rules}.
With this
decomposition, e.g. an incoming fermion with positive spin is given by
\begin{equation}
u^{+}(p)
    =\begin{pmatrix}
    \frac{m}{\sql p^\flat q \sqr }
    \raisebox{-0.2\height}{\includegraphics[scale=0.40]{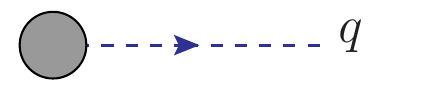}}  \\
    \phantom{\frac{m}{\sql p^\flat q \sqr}}
    \raisebox{-0.2\height}{\includegraphics[scale=0.40]{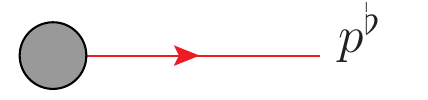}}
    \end{pmatrix}\;,
\end{equation}
which  reduces to a right-chiral Weyl spinor in the massless
limit $m \to 0$, $\pflat \to p$. Spinors for the other
external fermions are given in \appref{sec:additional rules}.

Vector bosons necessitate both a left- and a right-chiral spinor,
and can be represented by \cite{Berends:1981rb,DeCausmaecker:1981jtq, Mangano:1987xk, Gunion:1985bp, Gunion:1985vca,Kleiss:1986qc,Xu:1986xb,
Gastmans:1990xh},
\begin{align}
  \epsilon_{\blue{L}}(p_i,r) &
  = \frac{\rsqSp{i}\lanSp{r}}{\lan r i \ran}
  =\frac{1}{\lan r i \ran}
  \raisebox{-0.2\height}{\includegraphics[scale=0.35]{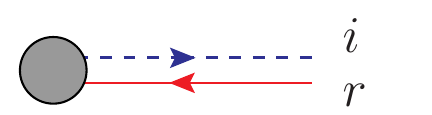}}
  \; \;
  \text{or}
  \; \; \, \,
  \frac{{\red|} r \rket \lbra i \blue{|}} {\rbra r i \rket}
  =\frac{1}{\lan r i \ran}
  \raisebox{-0.2\height}{\includegraphics[scale=0.35]{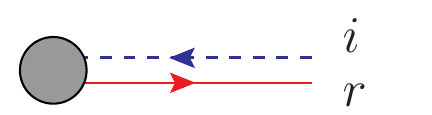}}\hspace*{-2 mm},\nonumber\\
  \epsilon_{\red{R}}(p_i,r) &
  =\frac{\rsqSp{r}\lanSp{i}}{\sql i r \sqr}
  =\frac{1}{\sql i r \sqr}
  \raisebox{-0.2\height}{\includegraphics[scale=0.35]{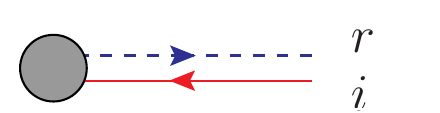}}
  \; \; 
  \text{or}
  \; \; \,\,
  \frac{{\red|} i \rket \lbra r \blue{|}} {\sql i r \sqr}
  =\frac{1}{\sql i r \sqr}
  \raisebox{-0.2\height}{\includegraphics[scale=0.35]{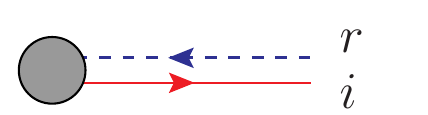}}\hspace*{-2 mm},
  \label{eq:pol vecs massless}
\end{align}
\commentMS{Modified, recheck}
with $\epsilon_{\blue{L}}$ corresponding to an incoming left helicity
state or outgoing right helicity state, and $\epsilon_{\red{R}}$
corresponding to an incoming right helicity state or outgoing left helicity state.
Note that the ``chirality'' ({\blue{L}}/{\red{R}}) of a gauge boson is given
by the spinor carrying its physical momentum $p_i$, and that $r$ is a gauge reference
momentum, with $p_i \cdot r \neq 0$ and $r^2 =0$.
The chirality-flow directions (arrows above) can be chosen arbitrarily, as long as the
two arrows making up the polarization vector are opposing each other,
and are consistent with the rest of the diagram, the
details of which we leave for \secref{sec:chiflowdirection}.
The unphysical reference vector $r$ 
is typically set to $r = (p^0, -\vec{p})$ both in \mga{} and in 
standard QFT literature (such as for example \citep{Peskin:1995ev}). We will
instead use this as a freedom to significantly speed up calculations,
by making diagrams vanish.

In principle these simplifications are not new to chirality flow
\cite{Mangano:1987xk}, but the flow picture makes it obvious how to
use the simplification even far inside Feynman diagrams.
To the authors' knowledge, the freedoms of choosing gauge reference
vectors as well as spin directions are unexploited in major event generators.
We discuss this subject further in \secref{sec:diagramremoval}, and explore the
benefits in \secref{sec:QCD}.

\subsection{Propagators and vertices}
\label{sec:internalchiflow}
In the previous section, we detailed the necessary structures to
describe free particles within the
chirality-flow formalism. To evaluate scattering amplitudes we
also need to represent propagators and vertices.

As there are two non-scalar particle types within 
the standard model, two types of propagators are needed. 
We first treat the gauge boson, which in the Feynman gauge has 
the structure of the Minkowski metric, translating to 
two opposing chirality-flow lines
\begin{align}
	\raisebox{-0.25\height}{\includegraphics[scale=0.4]{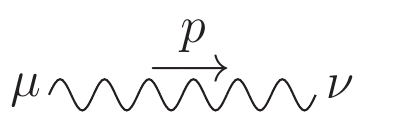}}
	&=  \quad  -i\frac{g_{\mu\nu}}{p^2}
	&
	&\rightarrow
	&
	&-\frac{i}{p^2}\raisebox{-0.25\height}{\includegraphics[scale=0.45]{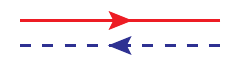}}
	\label{eq:vbpropagator}\;,
\end{align}
with directions to be matched with the rest of the Feynman diagram.

The fermion propagator has a more complicated structure, with a slashed momentum in
the numerator.
In the flow picture, this can be represented using the momentum-dot notation introduced in \citep{Lifson:2020pai},
\begin{eqnarray}
  \slashed{p} \equiv \sqrt{2}p^{\mu}\tau_{\mu}^{\da\be} = 
  \raisebox{-0.25\height}{\includegraphics[scale=0.4]{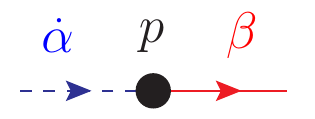}},
  \quad\quad
  \bar{\slashed{p}}\equiv\sqrt{2}p_{\mu}\taubar^{\mu}_{\al\db} =
  \raisebox{-0.25\height}{\includegraphics[scale=0.4]{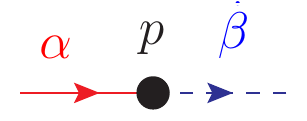}} \;,
\end{eqnarray}
\commentMS{inserted indices}
where we work in the chiral (Weyl) basis and use normalized versions of
the Pauli matrices, $\sqrt{2}\tau^{\mu} = \sigma^{\mu}$ and
$\sqrt{2}\taubar^{\mu} = \sibar^{\mu}$, to avoid factors of 2 in the Fierz identity.

The arrow direction aligned with the rest of the diagram must be chosen,
as elaborated on in \secref{sec:chiflowdirection}. 
For future purposes, we recall that these momentum-dots can be
rewritten in terms of outer products of Weyl spinors,
\begin{align}
\raisebox{-0.15\height}{\includegraphics[scale=0.4]{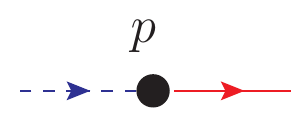}}~
&= \sum_i \sqrSp{i} \lanSp{i}
	& 
	&\text{and} 
	&
\raisebox{-0.15\height}{\includegraphics[scale=0.4]{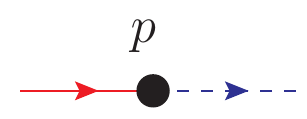}}~
&= \sum_i \ranSp{i} \sqlSp{i}
\label{eq:mom dot decomp}
\end{align}
\commentMS{Insereted, check}
for $p_i^2 = 0$ such that $p=\sum_i p_i$. 
The fermion propagator is then
\begin{align}
   \raisebox{-0.1\height}{\includegraphics[scale=0.4]{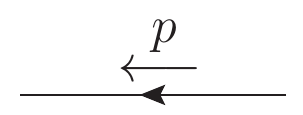}}
   &= \
\frac{i}{p^2-m^2} 
  \begin{pmatrix} 
    m {\delta^{\da}}_{\db} & \sqrt{2}p^{\da \be} \\ 
    \sqrt{2}\bar{p}_{\al \db} &  m {\delta_{\al}}^{\be}
  \end{pmatrix}
  = \frac{i}{p^2-m^2} 
  \begin{pmatrix} 
    m\includegraphics[scale=0.34]{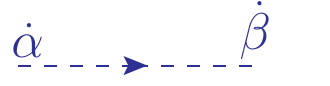}
    & \includegraphics[scale=0.34]{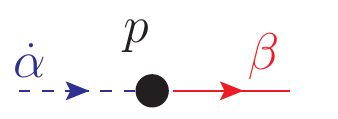} \\ 
    \includegraphics[scale=0.34]{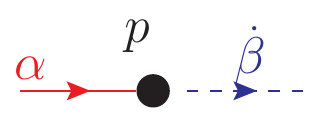}
    & m\includegraphics[scale=0.34]{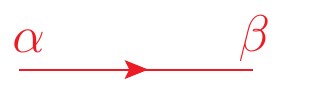}
  \end{pmatrix} \;.
  \label{eq:fermpropagator}
\end{align}

Finally, external particles and propagators are combined in QCD vertices. The simplest vertex is the fermion-vector vertex, represented by just a flow,
\begin{align}
	\includegraphics[scale=0.45,valign=c]{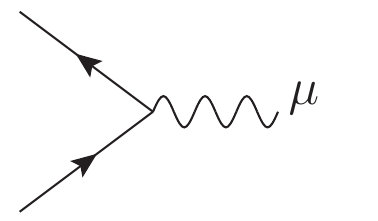}\!\!\!\!\! =
	\frac{ig_s t^a_{i \bar{j} }}{\sqrt{2}} \gamma^\mu
	&=
	\frac{ig_s t^a_{i \bar{j} }}{\sqrt{2}} \begin{pmatrix}
		0&   \sqrt{2}\tau^{\mu,\da\be}
		\\
		\sqrt{2}\taubar^{\mu}_{\al\db} & 0
	\end{pmatrix}
	\nonumber\\
    &\rightarrow
	ig_s t^a_{i \bar{j} } \begin{pmatrix}
		0&   \raisebox{-0.35\height}{\includegraphics[scale=0.35]{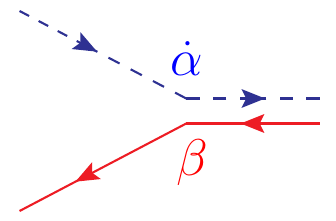}}
		\\
		\raisebox{-0.4\height}{\includegraphics[scale=0.35]{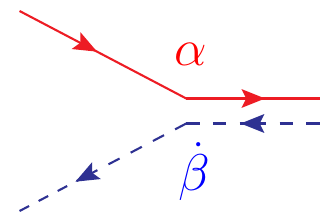}} & 0
	\end{pmatrix}~,
	\label{eq:fermion_gluon_vertex}  
\end{align}
with $g_s$ being the strong coupling constant and 
where the $\sqrt{2}$ in the denominator comes from the normalization
of our SU(3) generators $t^a_{i \bar{j}}$, which as our matrices $\tau$,
are normalized to one,
$\textrm{Tr}(t^a t^b) =\delta^{ab}$, $\textrm{Tr}(\tau^i \tau^j) =\delta^{ij}$.

The four-gluon vertex has a simple form in chirality flow, being represented by
Kronecker deltas that can be connected in three different
ways, 
\begin{align}
	\raisebox{-0.425\height}{\includegraphics[scale=0.275]{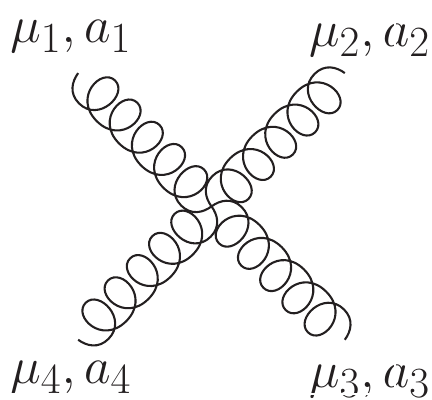}}
	&=\;\;
	i\left(\!\frac{g_s}{\sqrt{2}}\right)^{\!\!2}\!\!
	\;\sum\limits_{Z(2,3,4)}
	if^{a_1a_2b}if^{ba_3a_4}
	\;\Big(
	g^{\mu_1\mu_3}g^{\mu_2\mu_4}
	-g^{\mu_1\mu_4}g^{\mu_2\mu_3}
	\Big)\;
	\nonumber\\
	&\rightarrow \;\;
	i\left(\!\frac{g_s}{\sqrt{2}}\right)^{\!\!2}\!\!
	\;\sum\limits_{Z(2,3,4)}
	if^{a_1a_2b}if^{ba_3a_4}
	\left(\includegraphics[scale=0.225,valign=c]{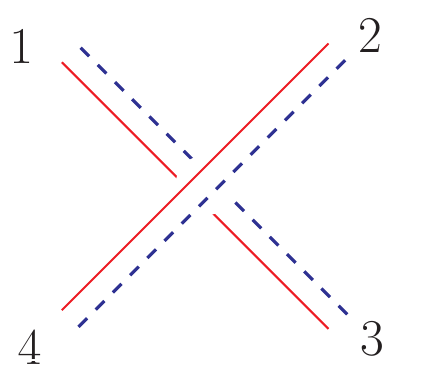}
	-
	\includegraphics[scale=0.225,valign=c]{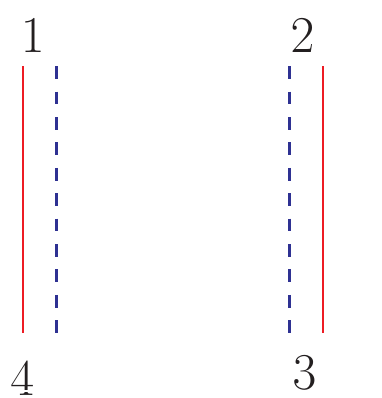}
	 \right)~,
	\label{eq:four_gluon_vertex}
\end{align}
where $Z(2,3,4)$ denotes the set of cyclic permutations of the integers
$2,3,$ and $4$, $f^{abc}$ are the SU(3) structure constants, and where arrow
directions which give flows aligned with the rest of the diagram
are chosen, as elaborated on in \secref{sec:chiflowdirection}, 

Finally, we address the triple-gluon vertex, which comes with the highest
complexity from the chirality-flow perspective as it involves
a momentum-dot. Chirality can flow between any two of the gluons (connected with
the metric), while the third gluon
is contracted with a momentum-dot,
\begin{align}
\includegraphics[scale=0.35,valign=c]{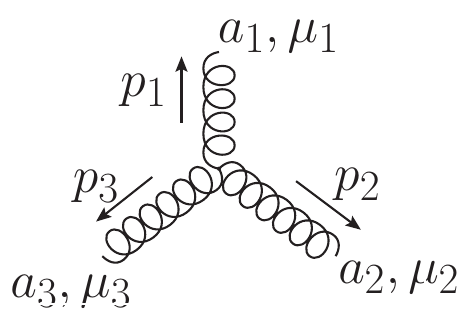}
\!\!\!
&=
\;
i\frac{g_s}{\sqrt{2}} 
if^{a_1a_2a_3}
	\Big(
	g^{\mu_1\mu_2}(p_1-p_2)^{\mu_3}
	+g^{\mu_2\mu_3}(p_2-p_3)^{\mu_1}
	+g^{\mu_3\mu_1}(p_3-p_1)^{\mu_2}
	\Big)
	\nonumber \\
&\rightarrow\;
i\frac{g_s}{\sqrt{2}} 
if^{a_1a_2a_3}\frac{1}{\sqrt{2}}
	\left(
\includegraphics[scale=0.325,valign=c]{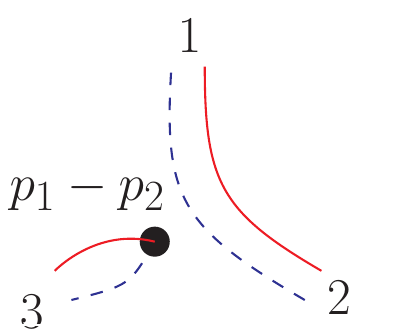} \!\!\!
+ \includegraphics[scale=0.325,valign=c]{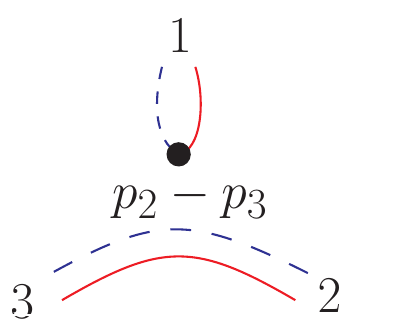} \!\!\!
+ \includegraphics[scale=0.325,valign=c]{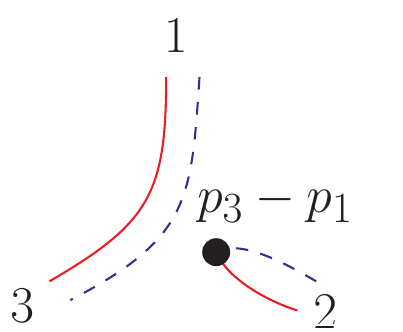}\!\!\!
\right)~,
\label{eq:trpgluvrt}
\end{align}
where the implicit arrows again should be chosen to give a continuous flow.

\subsection{Gauge based diagram removal and chirality-flow diagram removal}
\label{sec:diagramremoval}

In a recent paper we explored the benefits of gauge based diagram 
removal \citep{Lifson:2022ijv}, i.e, the freedom to choose reference 
momenta for external massless gauge bosons in such a way as to make a maximal number 
of Feynman diagrams vanish. There we worked with QED, and set the 
reference momenta $r_R$ of all right-chiral photons to be the physical 
momentum of a left-chiral lepton, and similarly for left-chiral 
photons with a right-chiral lepton.

As a simple example, before moving on to QCD, we consider the chiral structure of the process
$e^-_L e^+_R \to n\gamma$, using the chirality-flow part of the vertex in
\eqref{eq:fermion_gluon_vertex}, the massless fermion propagator from \eqref{eq:fermpropagator},
and massless external fermions and bosons. The corresponding diagram can then be
immediately written down as

\begin{eqnarray}
& &  \quad\quad\quad\quad\raisebox{-0.5\height}{\includegraphics[scale=0.5]{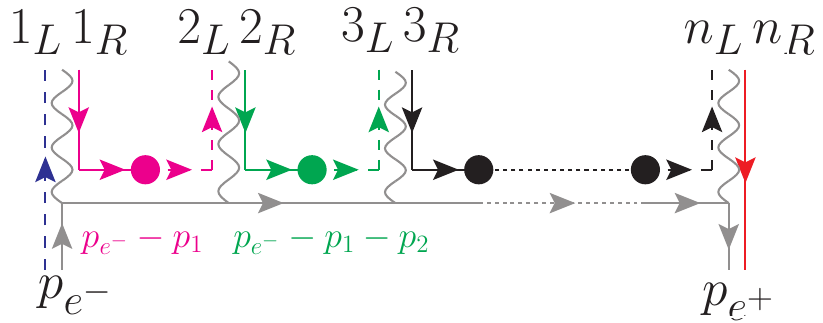}}
   \nonumber\\
  &\propto&  {\textcolor{blue}{[p_{e^-}\,1_L  ]}}
    \times {\textcolor{magenta}{
        \langle  1_R|\bar{\slashed{p}}_{e^- - 1}  |2_L  ] 
    }}\times 
   {\textcolor{ForestGreen}{
        \langle 2_R| \bar{\slashed{p}}_{e^- - 1 - 2}|3_L]}}
    \times \hdots \times 
           {\textcolor{red}{  \langle  n_R\,p_{e^+}  \rangle} }\;.
    \label{eq:qed_example}
\end{eqnarray}
In passing, we note that the chirality-flow arrows are chosen to align along the
fermion line, but to oppose each other for the photons, and that we could equally
well have chosen to swap all arrows since, employing \eqref{eq:mom dot decomp}, we have an even number of spinor
inner products.
As there is only one fermion line, each 
Feynman diagram can be described by the ordering of the photons.
However, by setting the reference momentum of right-chiral
photons to the momentum of the left-chiral fermion (and opposite),
\begin{align}
   i_L= r_R= p_{e^-_L} \quad \textrm{for right-chiral and} \quad i_R =r_L= p_{e^+_R} \quad \text{for left-chiral}, 
\end{align}
any diagram where a right-chiral photon is contracted with 
the left-chiral fermion will include a factor 
$\sqlSp{p_{e^-_L}} p_{e^-_L} \sqr = 0$ and vanish, 
and similarly for left-chiral photons.
It is therefore not possible to attach a right-chiral
photon next to $e^-_L$ or a left-chiral photon next to  $e^+_R$.
In principle this simplification is present in
the spinor-helicity formalism itself, but the chirality-flow formalism makes
it completely transparent.

As we consider non-abelian gauge theories, we encounter 
structures with several contributing chirality flows from the gauge boson
vertices. Similarly massive fermion propagators give rise to more
than one chirality flow. Here the choice of reference momenta or spin 
axes may not remove the full Feynman diagram despite removing some of 
the corresponding chirality-flow diagrams. 
Consider e.g. the chiral structure of the Feynman diagram
(now counting all momenta as outgoing)
\begin{eqnarray}
  & & \raisebox{-0.5\height}{\includegraphics[scale=0.4]{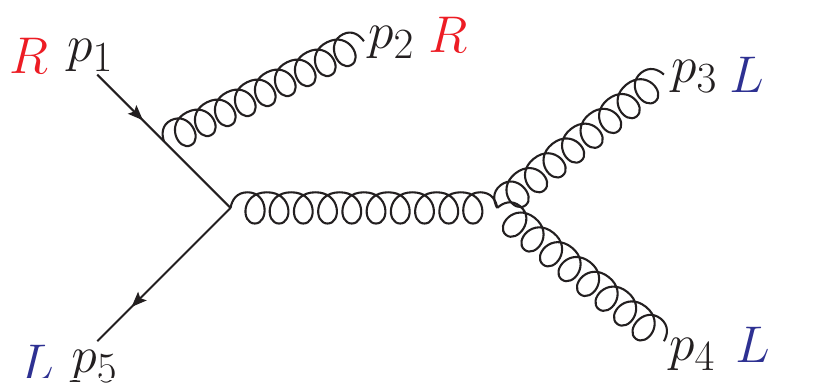}}\rightarrow
  \underbrace{\raisebox{-0.5\height}{\includegraphics[scale=0.4]{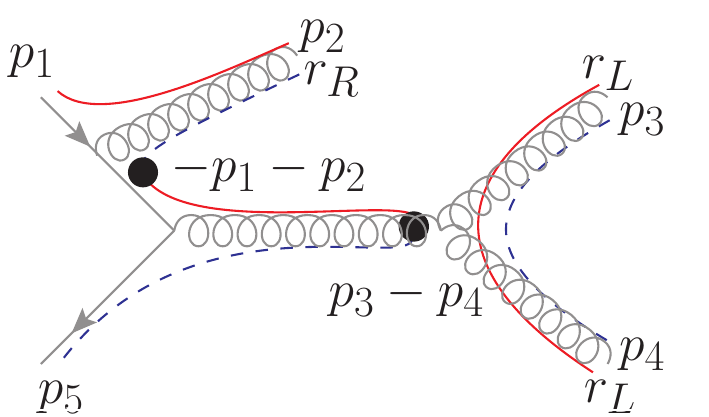}}}_{0}
  \label{eq:qcd_example} \\
  &&
  +\raisebox{-0.5\height}{\includegraphics[scale=0.4]{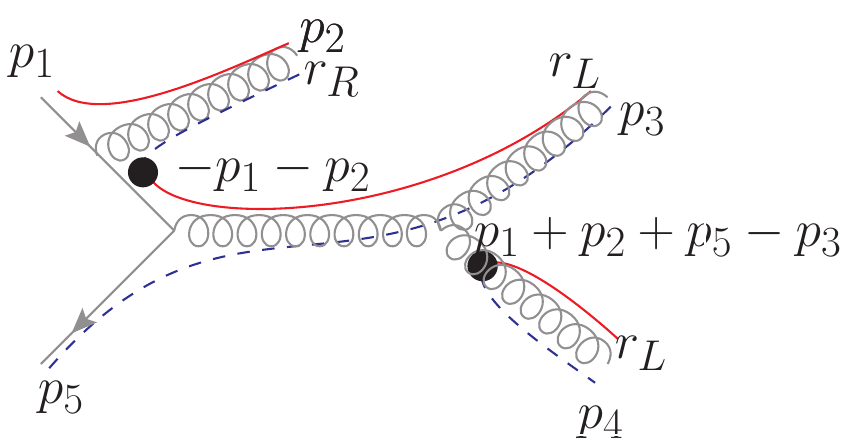}}
  +\raisebox{-0.5\height}{\includegraphics[scale=0.4]{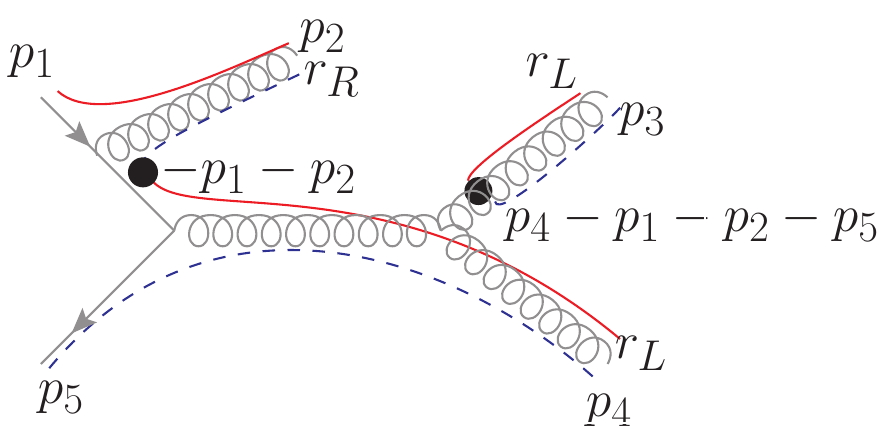}}\;,\;\nonumber 
\end{eqnarray}
\commentMS{Updated pictures fixed sign in momentum-dot}
which contributes to the process $q_{1,R} \bar{q}_{5,L} \to g_{2,R}\, g_{3,L}\, g_{4,L}$
for massless quarks.
Here we note that  setting the gauge reference vectors of the left-chiral gluons,
to the same momentum $r_3 = r_4 = r_L$, the first chirality-flow
diagram vanishes for any $r_L$.

Despite not having removed the total contribution from this Feynman diagram, 
we have reduced the complexity of the Lorentz structures, by removing some
chirality-flow diagrams. Therefore, we shall refer to this simplification
as \textit{gauge based chirality-flow diagram removal}, rather than 
gauge based Feynman diagram removal.

Further, if we let  $r_3 = r_4=r_L=p_1$, the gluons
$g_3$ and $g_4$ cannot attach to $q_1$, implying that all such Feynman diagrams vanish.
Similarly, letting  $r_2 = r_R = p_5$ gluon 2 cannot attach to $\bar{q}_{5}$.
This type of simplification is in direct analogy with the
gauge based Feynman diagram removal for QED as in \eqref{eq:qed_example}.
However, the 
simplification that arises from the contraction between external gauge bosons
is new for QCD.
This is elaborated upon  in \secref{sec:QCD}, where gauge specific Feynman
rules are introduced. Additionally, by considering sums
over measurements on general spin axes, rather than helicity sums, 
there will at times be gauge based chirality-flow diagram removal
in the fermion-vector vertex, where only one of the chiral 
components of a massive fermion contributes to the scattering amplitude.

\subsection{Chirality-flow direction}
\label{sec:chiflowdirection}
In the previous sections, we have largely neglected the question of
chirality-flow direction, simply stating that it has to be made
consistent.
With the examples in \eqref{eq:qed_example} and
\eqref{eq:qcd_example} 
at hand, we are now in a good position to comment on this.
To this end, we remind the reader that momentum-dots can be
written as sums of massless spinor outer products (cf. \eqref{eq:mom dot decomp}).

Each single chirality-flow line maps to a spinor inner product,
and considering massless fermions,
we note that the momentum-dot in the fermion propagator
ensures that every massless fermion line will have an even number of 
chirality-flow lines. Additionally, all vector bosons have two chirality-flow lines
with opposing flow direction. In massless QCD and QED,
there is thus always an \textit{even} number of flow lines,
making it impossible to flip an \textit{odd} number of spinor inner
products as long as flow continuity and the opposing arrow convention
for gauge bosons are respected.
By the antisymmetry of the inner product,
all consistent choices of flow directions then give the same amplitude. 
As an example, we may choose the arrow assignment 
\begin{equation}
  \raisebox{-0.5\height}{\includegraphics[scale=0.4]{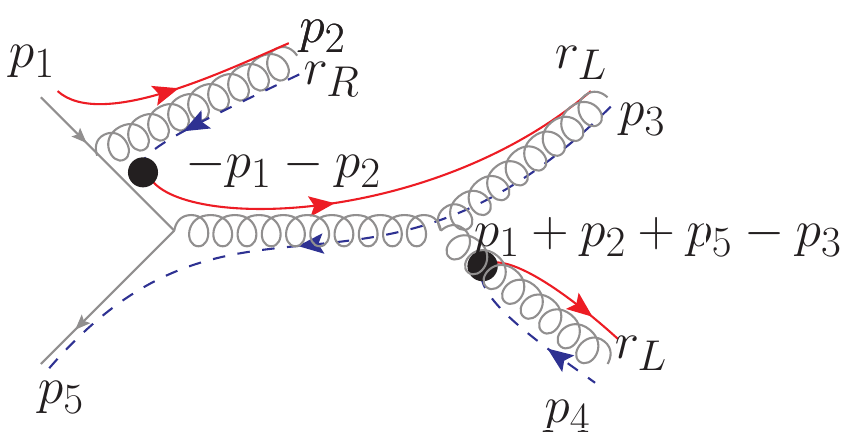}}
  \label{eq:arrows_exampel}
\end{equation}
for the second term in \eqref{eq:qcd_example}.

The massive fermion propagator, however, gives rise to terms with an
odd number of flows. It is then necessary to keep track of the
relative sign of the mass terms and the momentum terms.

\section{Flowing HELAS}
\label{sec:fhelas}

\begin{figure}[t]
 \begin{center}
   \raisebox{-0.5\height}{\includegraphics[scale=0.4]{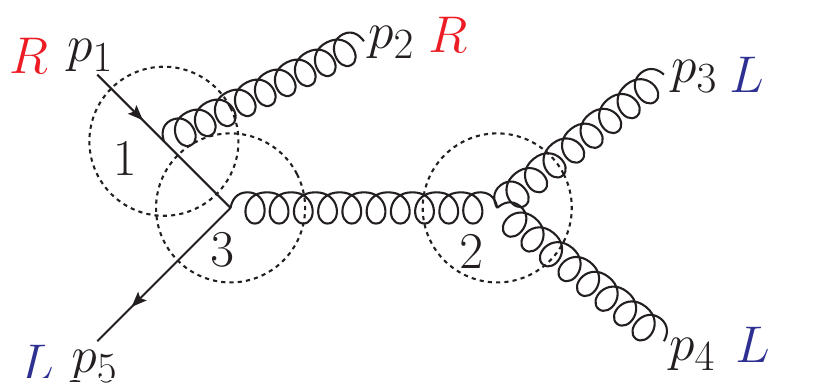}}
  \caption{
    \label{fig:helas}
    Example of the HELAS algorithm. Helicity amplitudes are calculated by
    iteratively combining particles to off-shell particles/wavefunctions.
    In the above case, particle 1 and 2 would be combined to a new off-shell fermion
    in vertex (1), and gluon 3 and 4 would be combined to one new particle in vertex (2).
    These two off-shell particles would merge with particle 5 in the last vertex (3).}
  \end{center}
\end{figure}

\mga{} evaluates helicity amplitudes using ALOHA-generated \cite{deAquino:2011ub} code based on 
a recursive algorithm as for HELAS \citep{Murayama:1992gi}.
The key principle of
this HELAS-like code is the recursive construction of propagating off-shell
particles (wavefunctions), as described in \figref{fig:helas}.

Every intermediate state (constructed
in a vertex combining two or three existing, on- or off-shell, particles) is numerically
evaluated as an off-shell wavefunction which feeds into the rest of
the calculation. Eventually three or four (on- or off-shell) particles
may join in a vertex, to create the amplitude corresponding to the
Feynman diagram in question. We will refer to this algorithm as the
HELAS algorithm.

In line with previous work \citep{Lifson:2022ijv} this structure 
is maintained in the current chirality-flow implementation, 
immediately enabling the same level of (off-shell) wavefunction
recycling as in standalone \mga{}, as well as a transparent speed
comparison, where it is ensured that any speedgain comes either
from the simplified Lorentz structures or from the use of
reference and spin vectors.
Consequently, there are only
two major changes made in our implementation:
\begin{itemize}
\item [(i)] Rather than exporting
  ALOHA-generated (HELAS-like) 
  code, subroutines are called from a
  rewritten flow-based version of the HELAS library,
  which makes use of the simplified chirality-flow Lorentz structures in
  the numerical evaluation of propagators and amplitudes.
\item [(ii)]  The diagram
  generation routine is modified to account for particles with
  definite chiral (or spin) states. In particular, chiral massless
  fermions are introduced, and external gauge bosons are
  supplied with a reference vector (cf.\,\eqref{eq:pol vecs massless}), 
  allowing for
  gauge based Feynman and chirality-flow
  diagram removal
  already at code generation time.
\end{itemize}

Compared to the default \mga{} implementation this implies simplifications which
in essence appear at two different levels:
\begin{itemize}
\item [(i)] \textbf{Simplified Lorentz structures}:
  Chirality flow
  makes subroutines smaller than the
  corresponding routines in e.g. HELAS. This applies in particular to vertices with chiral
  fermions, but even massive fermion vertices are less computationally demanding. 
\item  [(ii)]
  \textbf{Gauge based Feynman or chirality-flow diagram removal}:
  The numerical evaluation is simplified further by the choice of reference momenta
  and spin axes, as contributions may be known to vanish already at code generation time. This reduces
  the computation length, either directly at the Feynman diagram generation
  level (as some vertices and thereby Feynman diagrams may vanish),
  or at the level of chirality-flow diagrams (since chirality-flow diagrams
  which are known to evaluate to 0 can be ignored). 
  For example, the kinematic factor multiplying one of the color
  structures in the four-gluon vertex may vanish. 
  These simplifications are further described in \secref{sec:QCD},
  and they provide acceleration primarily to the gluon-only vertices and the
  massless fermion-vector vertex. 
\end{itemize}

\subsection{Arrow direction for the HELAS algorithm}
\label{sec:arrowdirection}

As discussed in \secref{sec:chiflowdirection}, the choice of flow direction
is simple for pen and paper calculations.
For automated algorithms, however, it is necessary to ensure that a consistent choice of
chirality flow can be set not just for a single chirality-flow
diagram, but for all possible chirality-flow diagrams, while working
at the level of single vertices. Hence, a
convention which can be translated to a simple algorithmic choice
must be used.

Instead of assigning chirality-flow direction for each diagram individually,
we therefore use the convention that all left-chiral spinors
are treated as having \textit{inflowing} chirality (i.e.\;they are given by a
left-chiral bra $\blue{\lbra|}$) whereas right-chiral spinors are considered
\textit{outflowing} (i.e. they are given by a right-chiral ket $\red{|\rket}$).
The same convention is applied to gauge bosons, but we
note that while external gauge bosons can be treated as direct products
of left- and right-chiral spinors (cf. \eqref{eq:pol vecs massless}),
internal propagators
require a description in terms of a matrix ($\sim M_{\a \db}$) due to the
presence of the triple-gluon vertex and contributions from
more than one term.
For simplicity, we therefore represent all gluons in this
form.

With these conventions, spinor inner products are always constructed from
spinors of the same form (i.e.\;a bra with a bra and a ket with a ket).
As an example, the second chirality-flow diagram in \eqref{eq:qcd_example} is 
mapped to the arrow assignment
\begin{equation}
  \raisebox{-0.5\height}{\includegraphics[scale=0.4]{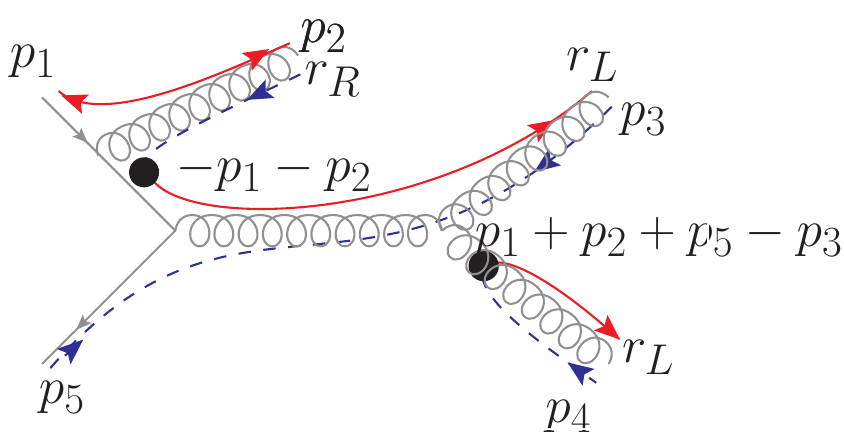}}\; \quad\text{(in code)}
  \label{eq:arrows_exampel_code}
\end{equation}
\commentMS{Check: had fermion prop sign mistake}
in our HELAS algorithm (\figref{fig:helas}).
Contractions then have to include systematic multiplication with
the Levi-Civita tensor to align the flows, i.e., always contract an
upper spinor index with a lower. Since the
ALOHA code is written in component form, this implies no speed loss.

\section{Gauge specific Feynman rules and simplifications from spin-direction}
\label{sec:QCD}

In this section we explore how knowledge of chirality and
spin can be used to define gauge dependent Feynman rules,
and how these can be exploited in the context of the
HELAS algorithm.
Although, clearly, the gluon reference vectors can be chosen independently for
each external gauge boson (for example, for a gluon with momentum
$(p^0,\vec{p})$, the custom choice is $(p^0,-\vec{p})$),
there are advantages with choosing them similarly for all
left- and right-chiral gluons respectively, as this gives rise to the
vanishing of more contractions.
We therefore limit our study to gauge choices where all external gluons 
of the same chirality have the same reference vector, which we further
take to equal some other spinor occurring in the process.

For internal gluons, the flow representation will make it
transparent how to deduce information about their spinor structure.
For example, the off-shell gluon arising in vertex 2 in \figref{fig:helas}
will necessarily come with the reference vector $r_L$ in
its right-chiral spinor, as seen from \eqref{eq:qcd_example}.
This means that we need to distinguish four different types
of gluons, left-chiral $g_L$, right-chiral $g_R$, general $g$ (i.e.\,no
specific knowledge of spinor structure),
and gluons which
are used for setting the reference momentum of
opposite-chirality gluons, $g_{LR}$ (or internal gluons which inherit this spinor structure).
These gluons
behave both as
left- and right-chiral gluons in the contractions.

A left-chiral gluon thus comes with a spinor structure
${\red|} r_L\rket  \lbra p_g{\blue|}$ in the code (generally alternatively 
 ${\blue |}p_g\lket  \rbra r_L{\red|}$),
and a right-chiral gluon comes with
${\red |}p_g \rket \lbra r_R {\blue |}$ in the code
(or generally
${\blue |}r_R \lket  \rbra p_g{\red |}$), whereas $g_{LR}$-gluons
come with
${\red |}r_L \rket \lbra p_g=r_R {\blue |}$ or
${\red |}p_g=r_L \rket \lbra r_R {\blue |}$ in the code
(or generally ${\blue |}p_g=r_R\lket \rbra r_L{\red|}$ or  
${\blue|}r_R \lket \rbra p_g=r_L{\red|}$).
Here we note that $p_g$ is the momentum of the gluon for an
external gluon, but is an ``inherited'' momentum entering as a
spinor argument for an internal gluon.
The inherited spinor structure can be exploited for subsequent
simplifications ``further in'' in the Feynman diagram, and we
will refer to gluons carrying the spinors
${\red |}r_L \rket$ and $\lbra r_R {\blue|}$
as \textit{effectively chiral}.

If we consider a process with only gluons,
it is clear that picking the reference momentum $r_L$
of the left-chiral gluons to equal the momentum of a right-chiral
gluon, additional spinor contractions vanish, implying additional simplifications.
For processes with both quarks and gluons it is less obvious if the
most advantageous choice is to set the reference vector after
a right-chiral gluon or after a right-chiral quark, something we discuss
further in \secref{sec:massless quarks}.

For massive particles, the corresponding degree of freedom becomes
physical, and is related to the direction in which spin is
measured, but we can still use the freedom to measure spin
in any direction to shorten the calculations.
We discuss this further in \secref{sec:spin and gauge Feyn rules}.
Beyond simplifications obvious from the chirality flow, we explore how the Schouten
identity can be used to further simplify the four-gluon vertex.

For the purpose of the HELAS algorithm, as discussed in \secref{sec:fhelas},
we need to distinguish between vertices resulting in off-shell particles
(wavefunctions) and vertices resulting in amplitudes.
At the first steps of the HELAS algorithm we have external
gluons with manifest helicities.
However, we will see that the internal (off-shell) 
gauge bosons often end up being effectively chiral.

\subsection{The triple-gluon vertex}
\label{sec:3gvertex}

We first consider the chiral structure of the three-gluon vertex,
\begin{align}
\includegraphics[scale=0.35,valign=c]{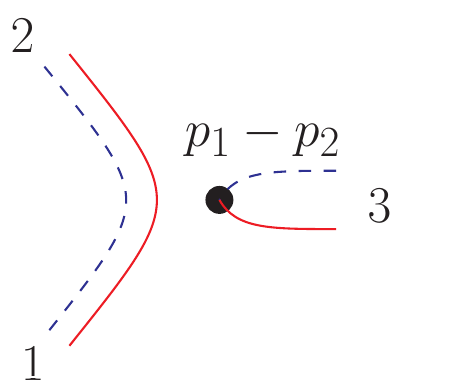} \!\!\! \quad
+ \includegraphics[scale=0.35,valign=c]{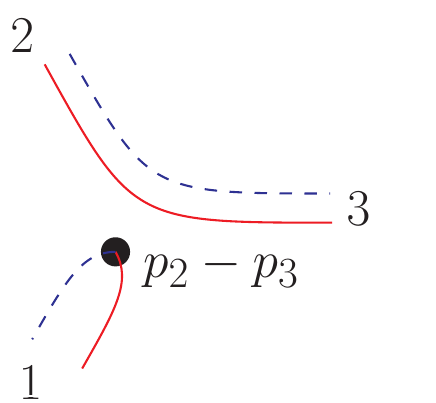} \!\!\!\quad
+ \includegraphics[scale=0.35,valign=c]{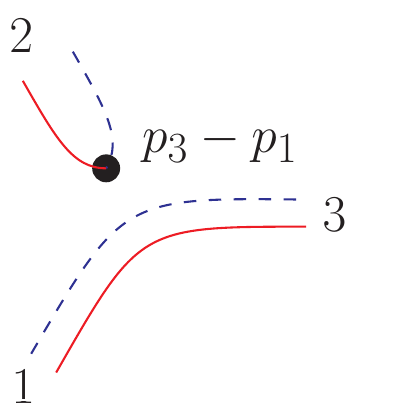}\!\!\! ,
\label{eq:3gGauge}	
\end{align}
assuming that all left-chiral gluons have the same reference momentum,
and similarly for all right-chiral gluons.
It is then immediately obvious that a vertex between three gluons
with equal chirality vanishes as the directly connected
double line gives  $\lbra r_{\red R} \textcolor{blue}{|} r_{\red R} \lket$ or $\rbra r_{\blue L} \textcolor{red}{|} r_{\blue L} \rket$
for all three terms, for example

\begin{align}
\underbrace{\includegraphics[scale=0.35,valign=c]{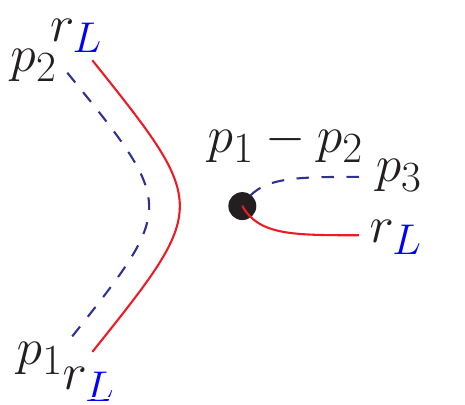}}_{ \propto \, \rbra r_{\blue{L}}  r_{\blue{L}}  \rket \, =  \,0} 
\!\!\! \quad
+\underbrace{\includegraphics[scale=0.35,valign=c]{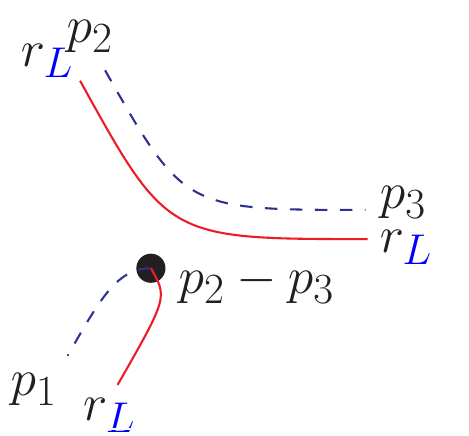}}_{ \propto \, \rbra r_{\blue{L}}  r_{\blue{L}}  \rket \, =  \,0} 
\!\!\!\quad
+\underbrace{\includegraphics[scale=0.35,valign=c]{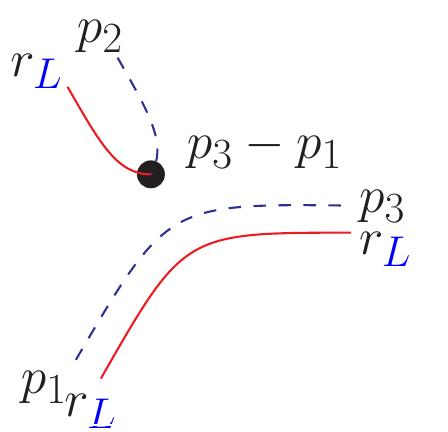}}_{ \propto \, \rbra r_{\blue{L}}  r_{\blue{L}}  \rket \, =  \,0} \;.
\label{eq:3gGaugeRRV}
\end{align}
Similarly, if two of the gluons share a known chirality, say
left, the term where the two left-chiral gluons are contracted
vanishes and only two terms remain.

Now consider vertices with $g_{LR}$. This
effectively left-right gluon will induce all of the simplifications for
left and right gluons above, for example in the 
$g_L \, g_R\,g_{LR} $ vertex, where with $g_1 = g_{L}$, $g_2 = g_R$, and
$g_3 = g_{LR}$ two out of three chiral structures are removed,
\begin{align}
\includegraphics[scale=0.35,valign=c]{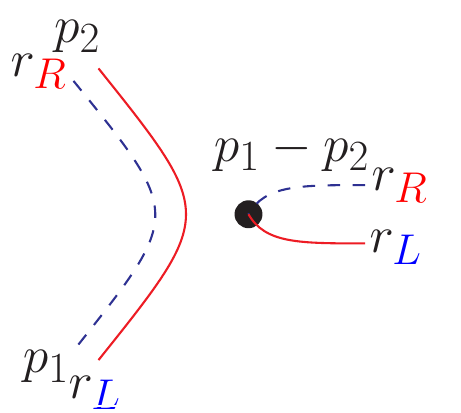} \!\!\! \quad
+ \underbrace{\includegraphics[scale=0.35,valign=c]{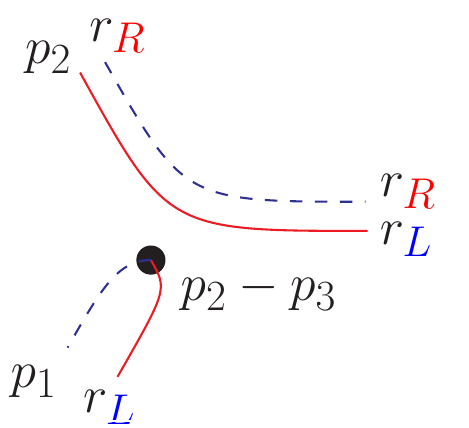}}_{\propto 
\, \sql r_{\red R} r_{\red R} \sqr \, = \, 0} \!\!\!\quad
+ \underbrace{\includegraphics[scale=0.35,valign=c]{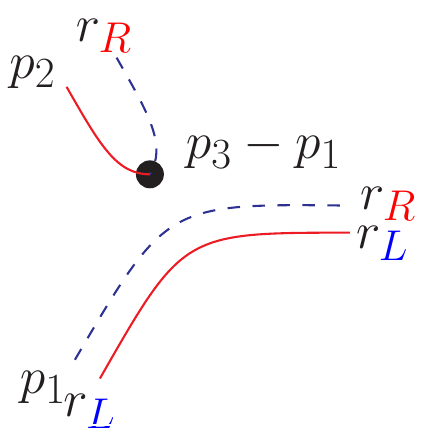}}_{\propto \, \lan r_{\blue L} r_{\blue L} \ran \, = \, 0} \!\!\! .
\label{eq:3gGaugeLRLR}	
\end{align}
Similarly a vertex between the left and right reference gluon
will always vanish since if say $g_1=g_L$, $g_2=g_R$
and $r_R=p_1$, $r_L=p_2$,
the first term vanishes immediately, whereas,
for external $g_1$ and $g_2$, 
the second
and third term vanishes by contractions with the momentum-dot using momentum conservation,
\begin{align}
\underbrace{\includegraphics[scale=0.35,valign=c]{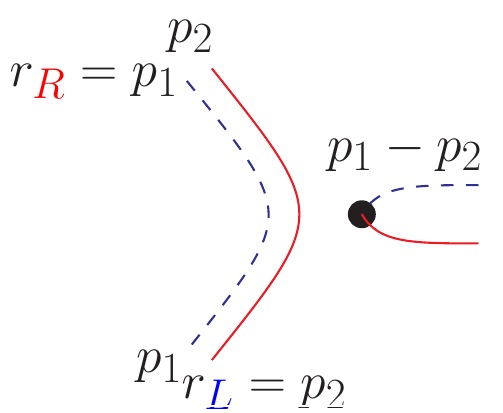}}_{ \propto \, \lbra p_1\,p_1\, \lket=0,\;\rbra p_2 p_2  \rket \, =  \,0} 
\!\!\! \quad
+\underbrace{\includegraphics[scale=0.35,valign=c]{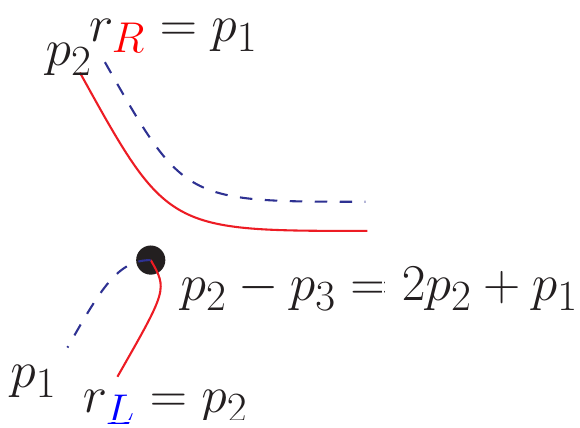}}_{0} 
\!\!\!\quad
+\underbrace{\includegraphics[scale=0.35,valign=c]{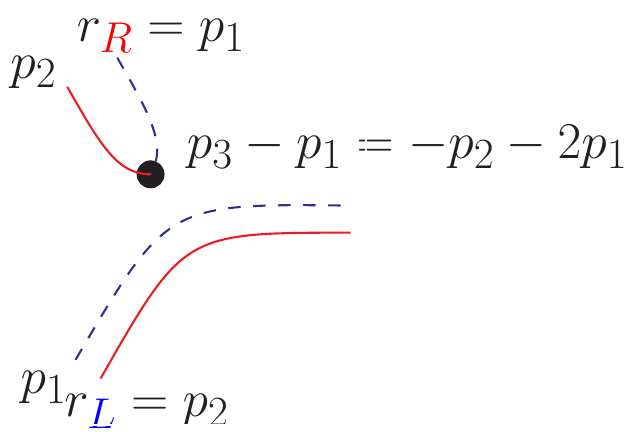}}_{0} \;,
\label{eq:3gGaugeRRV}
\end{align}
\commentMS{fixed mistake check again}
since
$p_2-p_3=2 p_2+p_1$ giving $2 \rbra p_2\, p_2\rket \lbra p_2 p_1\lket 
+ \rbra p_2\, p_1\rket \lbra p_1 p_1 \lket=0$, etc.
Therefore the external reference gluons can never couple directly to each other.

For the $g \, g_L \, g_R$-vertex, $g_L$ and $g_R$ contract to
each other as if they were unpolarized, implying that there is no gain to be found
in this case (unless one of them is a reference gluon, $g_{LR}$).
Similarly there is no simplification for vertices with no
or only one gluon with known chirality.
A systematic treatment of all versions of the triple-gluon vertex is given
in \appref{sec:vertex_table}.

\subsubsection{Vertices resulting in off-shell wavefunctions}

In the context of the HELAS algorithm, we note that the above 
treatment, with three gluons of (potentially) known chirality, is only applicable to
the last step, where three gluons result in an amplitude.
For the other steps, the created off-shell gluon will a priori
be of unknown chirality, and we rather want to 
\textit{derive} information about its chiral structure.
Thus we need to consider
two (on- or off-shell) gluons, $g_1$ and $g_2$ of (potentially)
known chirality, coming into a vertex
which results in a third off-shell gluon with derived
chiral structure.

If both $g_1$ and $g_2$ are left-chiral, we have
\begin{align}
\underbrace{\includegraphics[scale=0.35,valign=c]{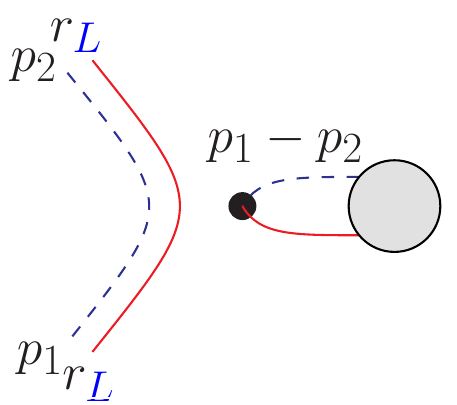}}_{ \propto \, \rbra r_{\blue{L}}  r_{\blue{L}}  \rket \, =  \,0} 
\!\!\! \quad
+ \includegraphics[scale=0.35,valign=c]{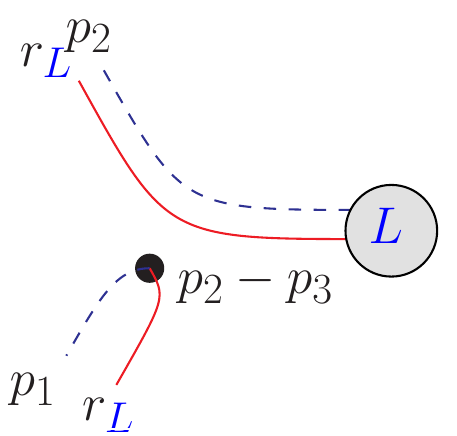} \!\!\!\quad
+ \includegraphics[scale=0.35,valign=c]{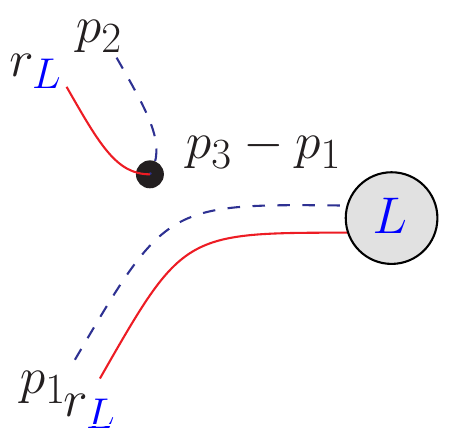}\!\!\! \quad,
\label{eq:3gGaugeRRV}
\end{align}
where the first term disappears due to the contraction between the (right-chiral) 
reference spinors for left-chiral gluons.
From the above spinor structure, we also deduce that the off-shell gluon
will effectively be left-chiral, i.e., when two gluons 
of the same chirality enter a three-gluon vertex, the resulting propagator will also 
act as having a well-defined chiral structure, which may continue to lead to further
simplifications further inside the Feynman diagram. For example,
the off-shell internal gluon resulting in vertex (2) in \figref{fig:helas}
will be effectively left-chiral.

Next, assume that $g_1$ and $g_2$ have opposite chirality, say $g_1$ left and $g_2$ right.
Then, generally, none of the terms in \eqref{eq:3gGauge} disappear, and the
resulting off-shell gluon can neither be treated as left nor right.
However, if we set e.g.\;the right-chiral reference momentum to be the momentum
of $g_1$ (which thus must be an external gluon), $r_R = p_1$, and thereby make
$g_1$ act as ``doubly chiral'' $g_{LR}$, we find
\begin{align}
\underbrace{\includegraphics[scale=0.35,valign=c]{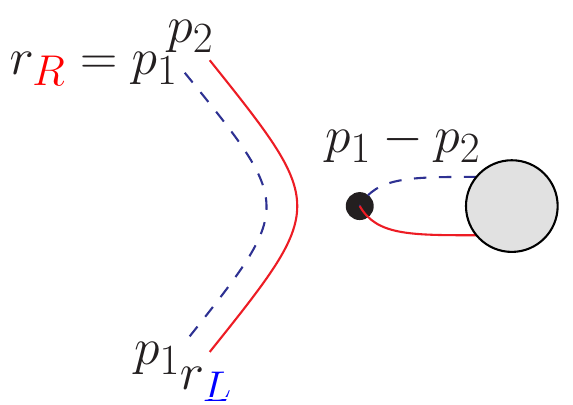}}_{ \propto \, \lbra p_1 \,p_1 \lket \, = \, 0} 
\!\!\! \quad
+ \underbrace{\includegraphics[scale=0.35,valign=c]{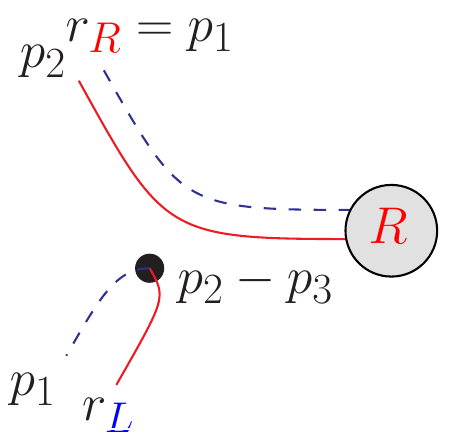}}_{\text{effectively right-chiral}}
 \!\!\!\quad
+ \underbrace{\includegraphics[scale=0.35,valign=c]{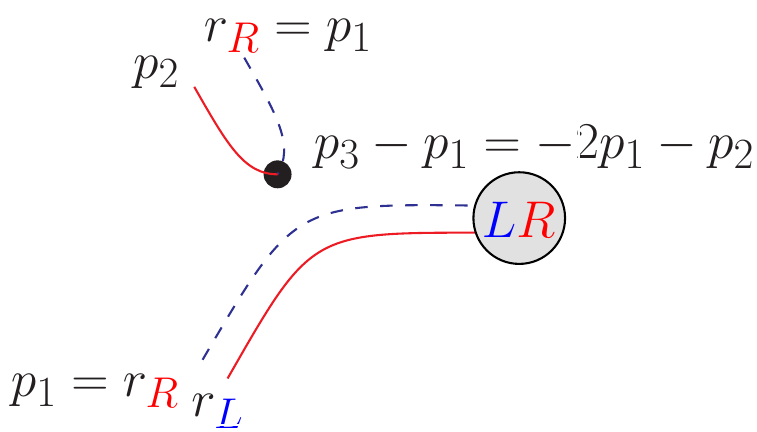}}_{ = \, 0,\text{ (if also $g_2$ external)}}\!\!\! ,
\label{eq:3gGauge1RL}
\end{align}
\commentMS{No mistakes, but changed on fig and propto. recheck}
where, for external $g_2$, the third term disappears due to contractions in the momentum-dot.
Regardless of whether $g_2$ is external or not,
the resulting gluon will effectively act as right-chiral in further vertices. 

If $r_R \ne p_1$, but $r_L=p_2$
in \eqref{eq:3gGauge1RL}, s.t. $g_2 = g_{LR}$,
the propagating gluon would instead act as left-chiral. Additionally, if $g_1 = g_{LR}$ 
and $g_2 = g_{LR}$, all three terms vanish by momentum 
conservation for external gluons.
A systematic treatment of all versions of the vertex is given
in \appref{sec:vertex_table}.

\subsection{The four-gluon vertex}

Next, we turn to the four-gluon vertex, which comes with three chiral structures
multiplying three different color structures
\begin{eqnarray}
	&&if^{a_1a_2b}if^{ba_3a_4}
	\left(\includegraphics[scale=0.225,valign=c]{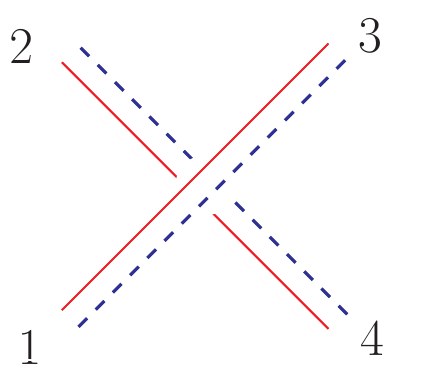}
	-
	\includegraphics[scale=0.225,valign=c]{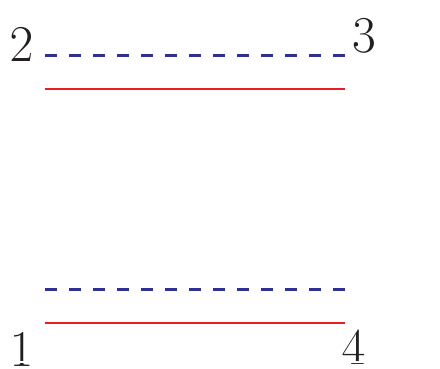}
	\right)\nonumber
        +
        if^{a_1a_3b}if^{ba_4a_2}
        \left(\includegraphics[scale=0.225,valign=c]{Jaxodraw/MS/4GluonVertexV2_Kin14}
	-
	\includegraphics[scale=0.225,valign=c]{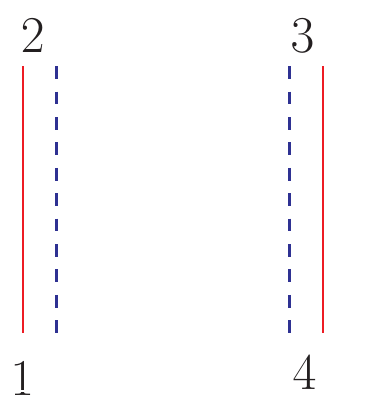}	
	\right)\nonumber\\
        &&+
        if^{a_1a_4b}if^{ba_2a_3}
	\left(\includegraphics[scale=0.225,valign=c]{Jaxodraw/MS/4GluonVertexV2_Kin12}
	-
	\includegraphics[scale=0.225,valign=c]{Jaxodraw/MS/4GluonVertexV2_Kin13}
	\right)~,
	\label{eq:four_gluon_expanded}
\end{eqnarray}
where we remark that the gluons which are contracted in color
are \textit{not} contracted in the chiral structure.

We first note that
this vertex vanishes fully whenever three (or four) gluons 
share the same chirality.
By considering \eqref{eq:four_gluon_expanded}, but taking two gluons to be
general, while keeping two gluons left (same chirality), we immediately see
that  one of the three spinor contractions vanishes.
For three or four general gluons, or for two gluons of opposite chirality,
no simplifications arise, whereas four-gluon vertices with $g_{LR}$
may result in many vanishing terms since these gluons can neither be
contracted with $g_L$ nor $g_R$. For example the vertex $g_{LR} g_L g_R g$
only has one surviving spinor contraction since $g_{LR}$ must
be contracted with $g$ for a non-vanishing result.

Additionally, we note that further simplifications can be obtained by exploiting
the Schouten identity, which in the flow picture can be represented as
\begin{align} 
\underbrace{\includegraphics[scale=0.45,valign=c]{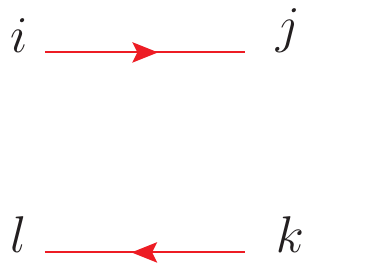}}_{
\lan ij\ran\lan kl\ran }
&=
\underbrace{\includegraphics[scale=0.45,valign=c]{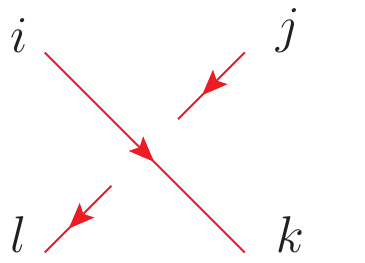}}_{
\lan ik\ran \lan jl\ran} 
+ 
\underbrace{\includegraphics[scale=0.45,valign=c]{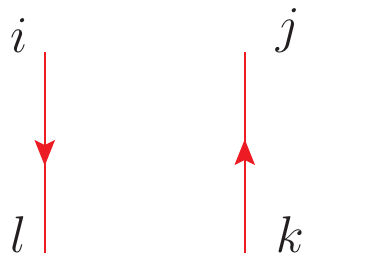}}_{
\lan il\ran\lan kj\ran }~,
\end{align}
and similarly for left-chiral spinors. Applying this to the four-gluon vertex yields 
simplifications beyond those immediately apparent from the flow. As an example, consider 
the $g_L g_L g_R g_R$-vertex (again with the particle positions from
\eqref{eq:four_gluon_expanded}), and apply the Schouten identity 
\begin{eqnarray}
	&&
	\includegraphics[scale=0.25,valign=c]{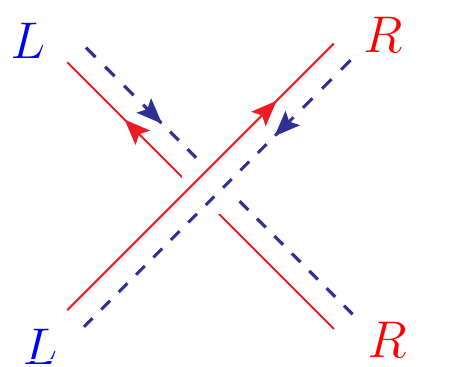}
	=
	-\underbrace{\includegraphics[scale=0.225,valign=c]{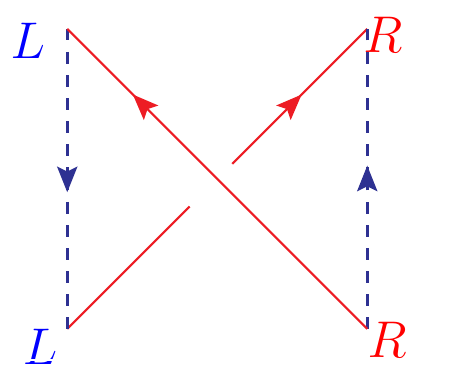}}_{\text{0}}
        +
        \includegraphics[scale=0.25,valign=c]{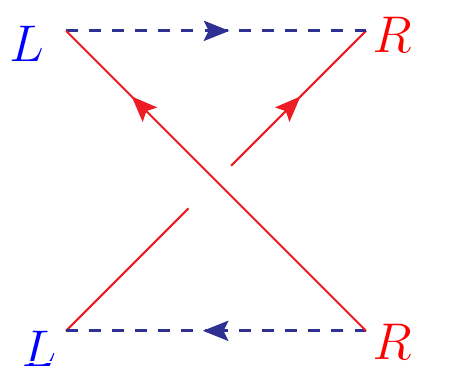}
        =-\underbrace{\includegraphics[scale=0.225,valign=c]{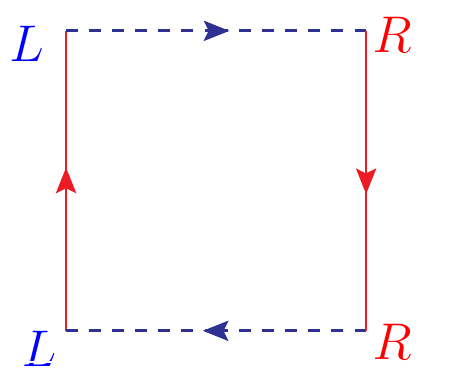}}_{0}
        +
        \includegraphics[scale=0.25,valign=c]{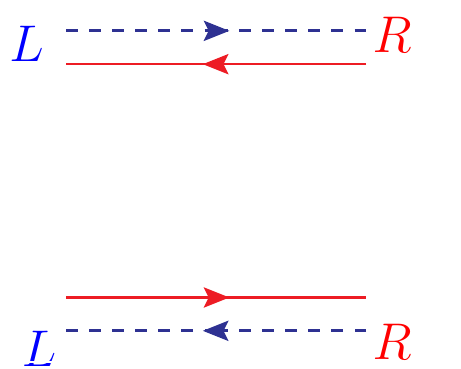}\;.
	\label{eq:four_gluon_LLRRSch}
\end{eqnarray}
\commentMS{FIxed red arrow}
Thus two of the chiral structures in \eqref{eq:four_gluon_expanded}
are equal and need only be evaluated once.
We therefore see that the chiral structures multiplying the first color structure
cancel by the Schouten identity. On top of that, the chirality flow with vertical lines
immediately vanishes,
implying that the two remaining color factors in
 \eqref{eq:four_gluon_expanded} multiply the same kinematic amplitude.

\subsubsection{Vertices resulting in off-shell wavefunctions}

We now consider vertices resulting in an off-shell gluon, assuming that $g_4$
will be the returned off-shell particle. Assume first that $g_1$
and $g_2$ have the same chirality, e.g $g_1$ and $g_2$ are both left-chiral whereas
$g_3$ is right-chiral. Then (keeping the same gluon positions as above) we have
\begin{eqnarray}
	&&if^{a_1a_2b}if^{ba_3a_4}
	\left(\underbrace{\includegraphics[scale=0.225,valign=c]{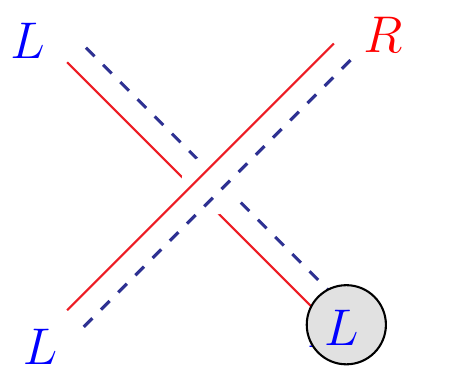}
	-
	\includegraphics[scale=0.225,valign=c]{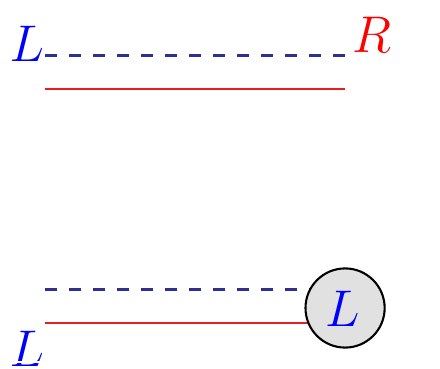}}_{\text{left}}
	\right)\nonumber
        +
        if^{a_1a_3b}if^{ba_4a_2}
        \left(
        \underbrace{\includegraphics[scale=0.225,valign=c]{Jaxodraw/MS/4GluonVertexOffLLR_Kin14}}_{\text{left}}
	-
	\underbrace{\includegraphics[scale=0.225,valign=c]{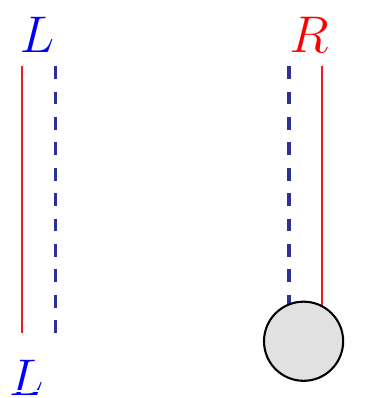}}_{0}	
	\right)\nonumber\\
        &&+
        if^{a_1a_4b}if^{ba_2a_3}
	\left(\underbrace{\includegraphics[scale=0.225,valign=c]{Jaxodraw/MS/4GluonVertexOffLLR_Kin12}}_{0}
	-
	\underbrace{\includegraphics[scale=0.225,valign=c]{Jaxodraw/MS/4GluonVertexOffLLR_Kin13}}_{\text{left}}
	\right)~,
	\label{eq:four_gluon_LLR}
\end{eqnarray}
i.e. the propagator resulting from a vertex with two gluons of the same (effective)
chirality will also have a definite effective chirality. Since external particles have
a well-defined chirality, ambiguity in the internal chiral states of gauge bosons arises
only from quark-gluon and three-gluon vertices.

We leave out the details here, but for vertices with only one chiral 
gluon or with two gluons of opposite chirality, no simplifications arise, whereas
further simplifications arise for the $g_{LR}$ gluons.
In total, identifying chiralities of internal gluons and applying Schouten identities allows 
for extensive gauge based (chirality-flow) diagram removal.
Again we leave out further details here, but provide the number of remaining kinematic terms 
for the different chiral configurations ($g_{LR}$, $g_L$, $g_R$ and $g$)
in \appref{sec:vertex_table}.

We also remark that the three spinor contractions multiplying
different color structures, have to be treated independently for further steps in the
HELAS algorithm.

\subsection{The fermion-gluon vertex}
\label{sec:fermion-gluon}

The final vertex we treat is the fermion-gauge boson vertex,
for which we consider the case of a massless and a massive fermion
separately.
For massless quarks, we always have a well-defined chiral state,
while for massive fermions we can explore the freedom to choose
spin direction in such a way that at least some chirality-flow
diagrams vanish.

\subsubsection{The massless fermion-gluon vertex}
\label{sec:massless quarks}
The (massless) fermion-gluon vertex has the structure of
\begin{eqnarray}
	&&
        \includegraphics[scale=0.4,valign=c]{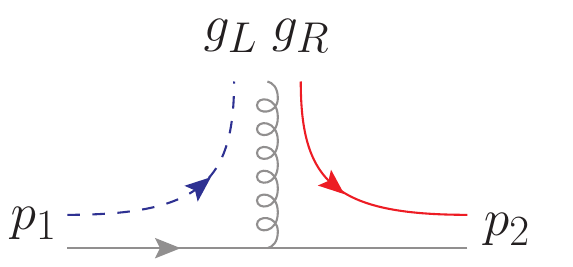}
	\label{eq:fermion_gluon}
\end{eqnarray}
from which it is immediately clear that if one of the spinors of the
gluon matches one of the spinors of the quark, the vertex will
vanish. This will happen for example if the gluon is right-chiral and
has the quark momentum $p_1$ as reference momentum, $\lbra p_1 \textcolor{blue}{|} g_L=r_R =p_1\lket=0$.
This is in direct analogy to QED  in \eqref{eq:qed_example}, where right-chiral photons
can not attach to the left-chiral reference 
fermion.
While we could alternatively set the gluon reference momenta according
to the momenta of other gluons, setting them to the momenta
of quarks turns out to give the best speedgain for high
multiplicities, as some vertices --- and thereby Feynman diagrams --- 
can be fully removed, rather than (typically) just be simplified,
as in the gluon-only case.\\

\hspace*{-0.8 cm}
\textbf{Vertices resulting in off-shell wavefunctions.}
As for the gluon vertices, we need to distinguish vertices resulting in
off-shell wavefunctions from vertices resulting in amplitudes for the purpose of the
HELAS algorithm.
If the vertex results in an off-shell fermion we have one of the
situations
\begin{eqnarray}
	&&
	\includegraphics[scale=0.4,valign=c]{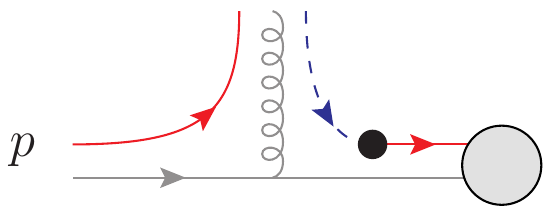} \quad \text{or} \quad
        \includegraphics[scale=0.4,valign=c]{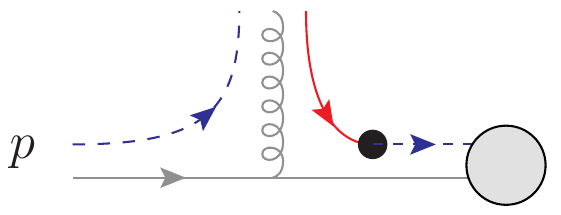}\;,
	\label{eq:four_gluon_LLRRSch}
\end{eqnarray}
where we also display the momentum-dot coming from the massless fermion
propagator. 

We note that if the gluon and the external quark
share a spinor, the contraction vanishes. Again this will only
happen if the gluon has the quark momentum as reference spinor.
If the diagram does not vanish, the fermion
continues in to a fermion propagator containing a momentum-dot.
The end of that propagator will then again have an explicit chirality
(the same chirality as the original incoming quark), 
while entering the \textit{next} vertex (gray blob above).
The quark thus remains chiral.
Technically this momentum-dot is included in the vertex implementation.

If instead, the fermions 
give an off-shell gluon,
we note that the gluon
can be effectively chiral if one of the fermion momenta is used
for the gluon reference momentum.

\subsubsection{Massive fermions with general spin direction}
\label{sec:spin and gauge Feyn rules}

As discussed in \secref{sec:spinors} we may explore the freedom of choosing the spin
axis $s^\mu$ to our benefit, such that some chirality-flow diagrams vanish.
We recall that the spin operator is defined as\footnote{
Note that the spin operator $\Si^\mu$ 
is directly related to the Pauli-Lubanski operator $W^\mu$ 
via $\Si^\mu = \frac{2}{m}W^\mu$.
We remind that 
$W^2 = -m^2J(J+1)$,
where $J$ is the total spin,
is one of the two quadratic Casimir operators of the Poincar\'e algebra
(with the other being $P^2=m^2$).} (see e.g.\ \cite{Ohlsson:2011zz}),
\begin{equation} \label{eq:Sigma 4vec defn}
\frac{1}{2}\Si^\mu = -\frac{1}{4m}\eps^{\mu\nu\la\omega}P_\nu\si_{\la\omega}~,
\end{equation}
where $P_{\nu}$ is the momentum operator ($P_{\nu} = i\partial/\partial x^{\nu}$),
$\eps^{\mu\nu\la\omega}$ ($\eps^{0123}\defequal 1 \Rightarrow \eps_{0123}= -1$) 
is the four-dimensional Levi-Civita tensor,
and $\si^{\mu\nu}$ (giving the Lorentz generators for the $(1/2,1/2)$-representation) is defined as
\begin{equation} \label{eq:sigma mu nu}
\si^{\mu\nu} = \frac{i}{2}\left[\ga^\mu,\ga^\nu\right]~.
\end{equation}
The spin projected onto $\smu$ is then given by the operator
\begin{equation} \label{eq:spin on motion rest}
 \mathcal{O}_s=-\frac{\Si^\mu \smuLow}{2} =
 \frac{1}{4m}\eps^{\mu\nu\la\omega}\smuLow P_\nu\si_{\la\omega}~,
\end{equation}
such that the spin direction, in this sense, is actually only defined up to a four-momentum proportional
to the particle's momentum $p$. Adding any linear combination of $p$ to
$s = (1/m)\left( p - 2\alpha q \right)$ (\eqref{eq:spin_vec})
will result in the same $\mathcal{O}_s$\footnote{In the non-relativistic limit
$p=(m,0,0,0)$ and $s=(0,\hat{s})$.
Clearly, adding any linear combination of $p$ to $s$ or $q$ will leave
$\mathcal{O}_s$ invariant.}.
Therefore, we could equally well have used
$s' = -2\alpha q/m
=-m q/(p\cdot q)
$ for the spin direction.\footnote{This applies to $\mathcal{O}_s$ when expressed as
above, when rewritten as
$\mathcal{O}_s = \frac{1}{2}\ga^5\smu\ga_\mu~$
  the spin vector must be taken to be $s$.} 
\commentMS{New: check}
Thus $q$
plays the role of defining the \textit{other} four-vector (aside from $p$) which determines
the operator $\mathcal{O}_s$, and thereby what we mean with positive
and negative spin.

This we will use to our advantage. While the direction
$s = (1/m)\left( p - 2\alpha q \right)$ clearly depends on
the particle's momentum, the momentum $q$ can be taken to
be the same for all particles.  
For further simplifications, we may 
choose it to be the same as
the momentum for some other spinor in the process, such as for example
the reference vector of left-chiral gluons.

\subsubsection{The massive fermion-gluon vertex}

The structure of the massive fermion vertex (itself) contains
the flows
\begin{eqnarray}
	\includegraphics[scale=0.4,valign=c]{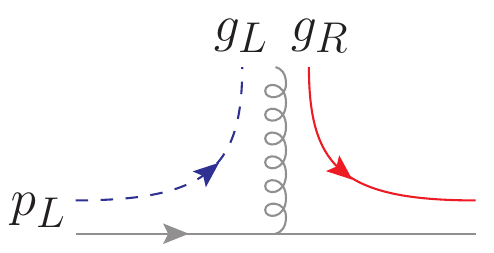} \; \text{and} \quad
        \includegraphics[scale=0.4,valign=c]{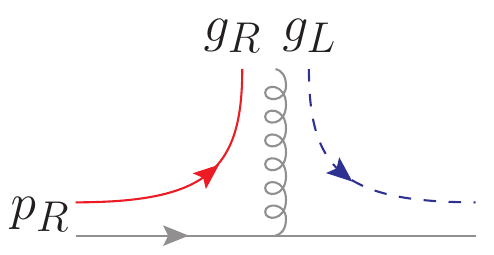} 
        \label{eq:qgOff}
\end{eqnarray}
for some spinors, for external fermions either
$\lbra p_L{\blue|}= \lbra q{\blue|}$ and $\rbra p_R{\red|}= \rbra p^\flat{\red|}$ 
or
$\lbra p_L{\blue|}= \lbra p^\flat{\blue|}$ and $\rbra p_R{\red|}= \rbra q{\red|}$,
as given in \eqsrefa{eq:ubarvpm_spin_cf}{eq:uvbarpm_spin_cf} in \appref{sec:additional rules}
(and accompanied by the phases and factors stated there).

As in \eqref{eq:fermion_gluon}, we may thus remove vertices with
for example $\lbra p_L=q=r_R \textcolor{blue}{|} g_L=r_R\lket$ if $p_L$ is set to equal the reference
spinor of the right-chiral gluons.
This momentum ($r_R$) may in turn be taken to equal the physical momentum
of a left-chiral gluon. 

Alternatively the reference vector $r_R$  of all
right-chiral gluons may be put to equal the
$p^\flat$ vector of some quark.
For a low number of gluons this will lead to many vanishing
chirality-flow diagrams, since the reference quark can not couple
to one of the gluons.
However the number of
terms that can be removed in this way scales badly with the number
of gluons. Overall, we have rather found it advantageous to
let the physical momentum of an opposite-chirality gluon dictate
the vector $q$ for the quarks, with $q=r_L/r_R=p_\text{some right/left gluon}$. 

With this choice, the full vertex will
never vanish, but for processes with many gluons, it turns out to give
the best overall speedup, due to the simplifications in the gluon
vertices.
We remark that this is contrary to the situation with massless
quarks, where the best high-multiplicity speedup is found by setting
the reference momenta to be the momentum of a quark (with opposite chirality).
From this, it is clear that (for example) if the left gluon spinor equals
${\blue|}p_L\lket$ the first part of the vertex disappears, but the
second term remains.

\hspace*{-0.8 cm} \textbf{Vertices resulting in off-shell wavefunctions.}
In general, for a vertex resulting in an off-shell fermion wavefunction,
we have terms of the form
\begin{eqnarray}
	&&
	\includegraphics[scale=0.4,valign=c]{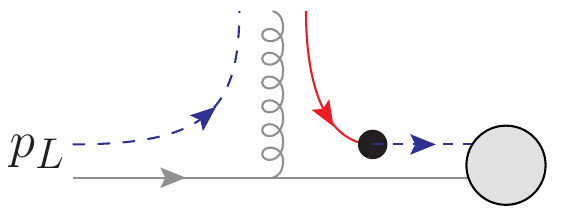} \; \raisebox{-5.5pt}{,} \; \phantom{\raisebox{-4.5pt}{\text{and}}} \quad
        \includegraphics[scale=0.4,valign=c]{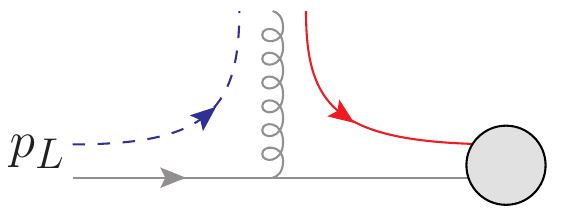} \, \raisebox{-5.5pt}{,} \quad \; \nonumber \\
       && \includegraphics[scale=0.4,valign=c]{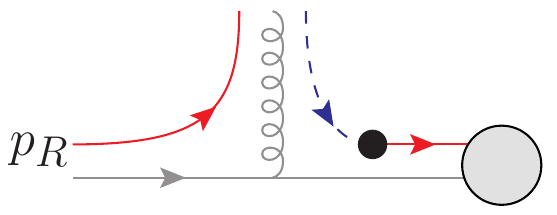} \; \phantom{\raisebox{-5.5pt}{,}} \; \raisebox{-4.5pt}{\text{and}} \quad
        \includegraphics[scale=0.4,valign=c]{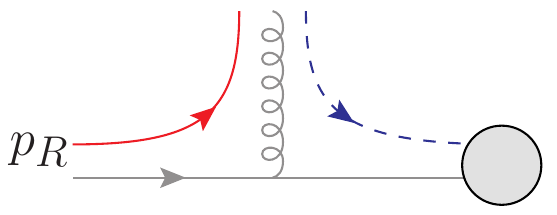} \; \raisebox{-5.5pt}{,} \quad
        \label{eq:qgOff}
\end{eqnarray}
where, like in the massless case, we choose to include the chiral
structure of the propagator, to get the spinor structure entering
the next vertex.

In general all four such terms may survive, but again, if the gluon
spinor contracted with the incoming quark is given by the same
momentum $p_L$ or $p_R$, two of the terms above will vanish,
which is used to build faster versions of the code. 
However, due to the massive propagator, the simple structure of one
outgoing helicity is nevertheless lost for the fermion entering
the next vertex.
In spite of this, 
our representation of the
fermion-gauge boson
vertices are less computationally demanding.

Finally, we note that the vertex may alternatively result in an off-shell gluon,
which then is of general type $g$.

\section{Results}
\label{sec:results}

In this section, we present time measurements for amplitude evaluations
in the chirality-flow implementation, and comparisons to 
the standalone version of \mga.
Since we use the same diagram generation and code generation 
routines as the native \mga\; framework, and only
(i)  speed up numerical subroutines
using a flow-based library, and (ii) remove (chirality-flow) diagrams
known not to contribute, we are confident that any improvement in
performance can be attributed entirely to chirality flow.
Since we only address the Lorentz structure, the squaring of the
color structure has been turned off for both versions of the code,
to assess the speedup of the Lorentz structure. This would otherwise be
a bottleneck for high multiplicities.

We first explore the speedup in the pure Yang-Mills case. Then,
bearing in mind that almost all fermions  are effectively massless
at LHC energies, we consider massless up quarks. Finally we explore
gluon induced $t\bar t$ production in association with up to three additional gluons.
In all cases we find an increasing speedgain with increasing multiplicity.

Time measurements were made on a MacBook Pro laptop with an Intel Core i7-4980HQ CPU
and runtime measurements are taken over 100 000 randomly generated 
phase space points. 
At present, our code generates each
chirality/spin configuration as a different runtime process, and phase space
points are hence generated with RAMBO \cite{Kleiss:1985gy} 
separately for each spin configuration for chirality flow, 
whereas this is done outside the spin sum for \mga.
For high multiplicities, the runtime for RAMBO is negligible, but
for small multiplicities this overhead slows down our implementation.
Even so, we find for all processes that the chirality-flow implementation outperforms
\mga\; already for five external particles.

It should be noted that our results are from the standalone
version. For helicity sampling, we expect them
to carry over to the full version, but it should be noted
that since recently \mga\; comes with helicity recycling,
where parts of Feynman diagrams are recycled between processes
with different helicities \cite{Mattelaer:2021xdr}. This can be
applied to our version as well. However some adjustment would be needed,
since to fully exploit helicity recycling, one needs to use the
same reference vectors ($r_L$ and $r_R$) for all helicity configurations, whereas
we have chosen them differently. Picking different $r_L/r_R$ for
different helicities will result in some speed loss, although
the bulk of the improvement will remain.

\subsection{Gluon processes}
\label{sec:gluon amps}

\begin{figure}[t]
\begin{center}
  \includegraphics[scale=0.35]{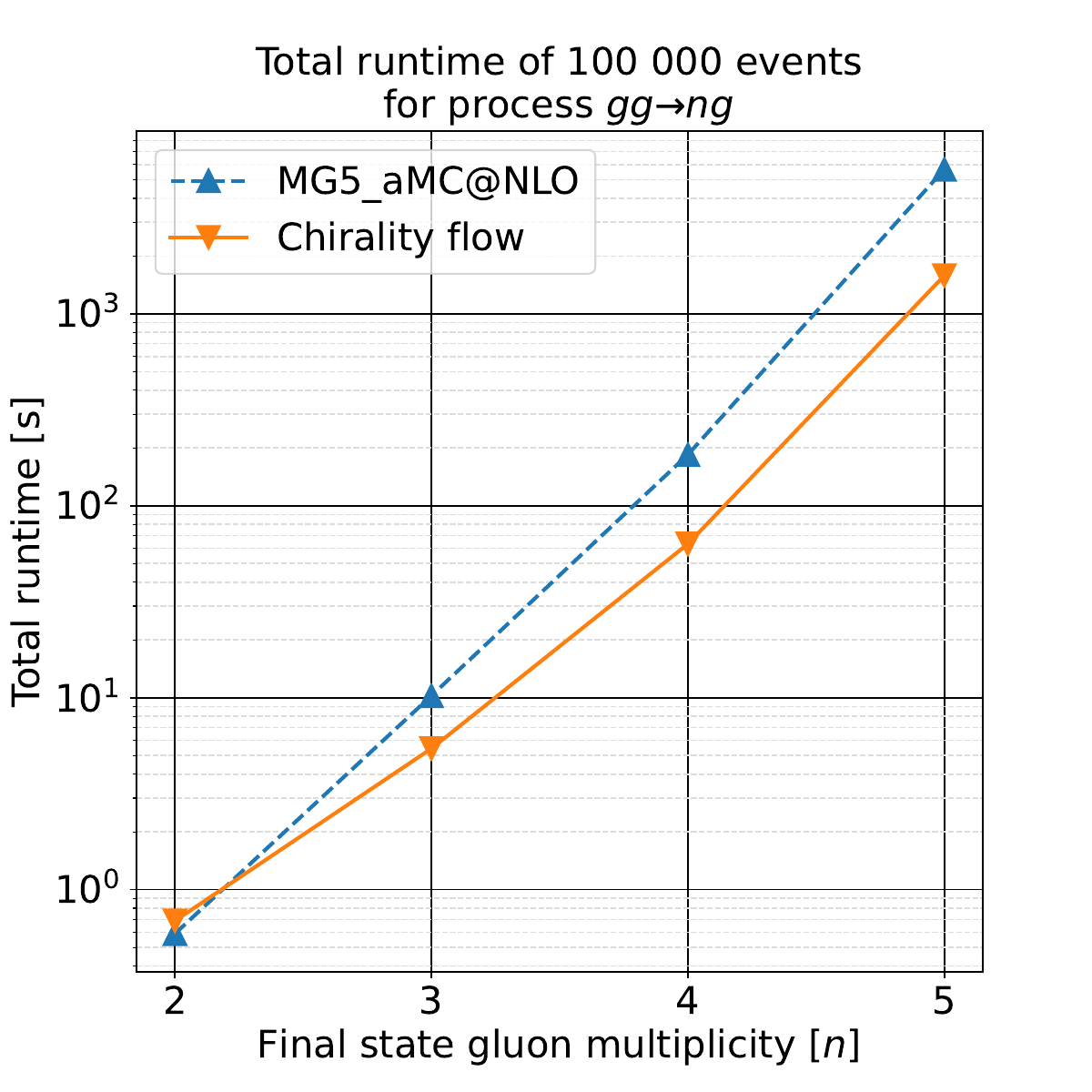}
  \includegraphics[scale=0.35]{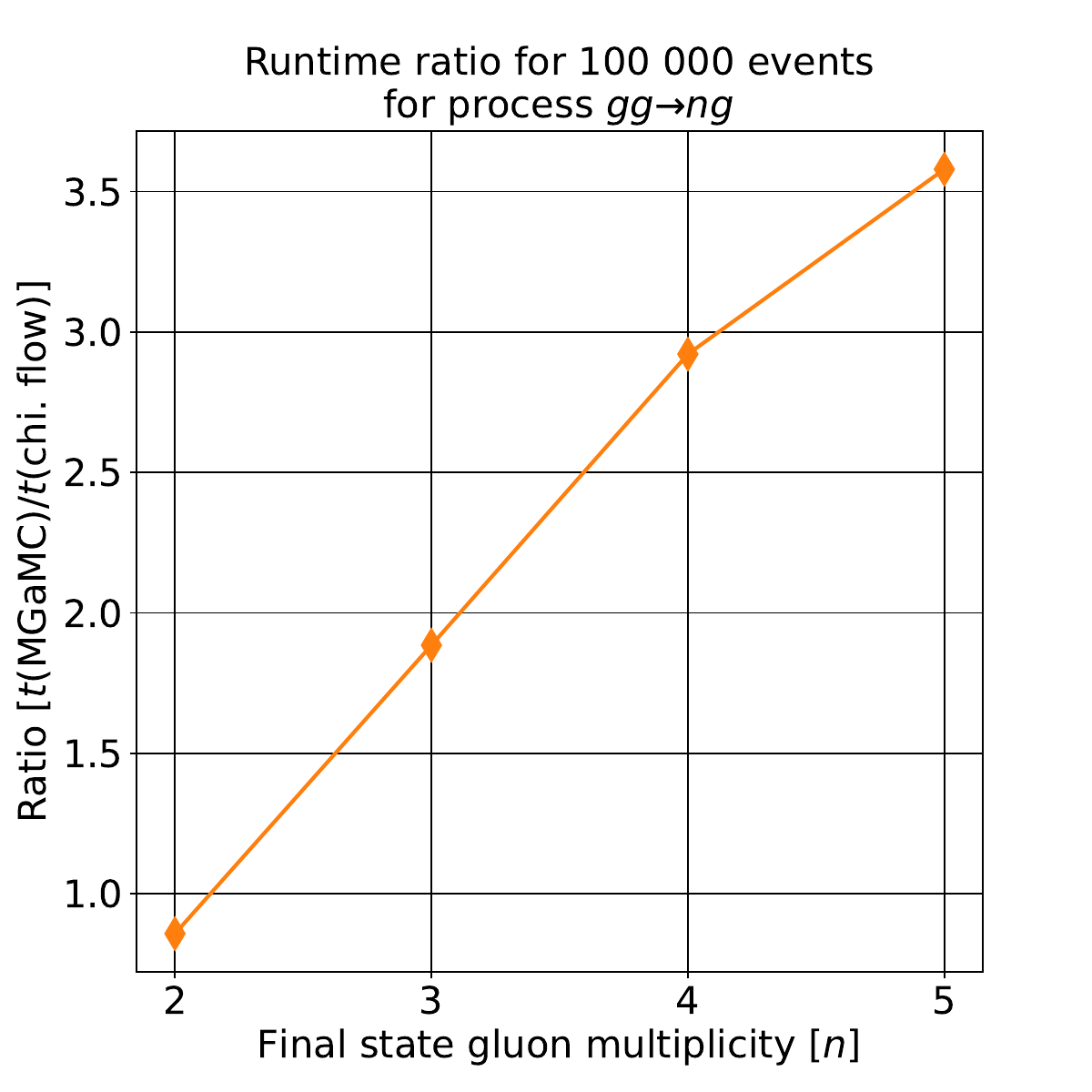}
  \caption{Runtimes (left) and ratio (right) for $gg\rightarrow n g$ for
    standalone \textsc{MadGraph5\_aMC@NLO}
    and for the chirality-flow branch of the code.
    The squaring of the color structure
    has been turned off to compare only the Lorentz
    structure treatment. In the current chirality-flow implementation,
    the RAMBO momentum sampling is done separately for each spin configuration,
    and this overhead has a runtime impact for low multiplicities.
    The amplitude evaluations themselves are consistently faster for chirality flow,
    but this code artifact outweighs the acceleration at multiplicity four.
    For high multiplicities, the phase space sampling takes a negligible fraction of
    the total runtime.}
\label{fig:gg_to_ng}
\end{center}
\end{figure}

We here explore the benefits of chirality flow in the pure gluon case,
$gg\rightarrow ng$, $n=2\hdots 5$.
Our reference vectors are set as discussed  
in \secref{sec:QCD},
i.e., left-chiral gluons with a reference vector for the right
spinor are treated as $\rbar r_L\rket\lbra p_g \lbar$, where the
reference vector $r_L$ is taken to be
the same for all left gluons. Further, it is taken to equal the
momentum of a right-chiral gluon (the right-chiral gluon 
with lowest number in \mga).
Reference vectors for right-chiral gluons are set similarly.

Runtime comparisons are shown in \figref{fig:gg_to_ng}.
We note that there is an increased speedup for high multiplicity.
This speedgain can be attributed to a high number of vanishing
chirality-flow diagrams when the vanishing of
$\lbra r_R \lbar r_R \lket$, $\lbra p_\text{ref}=r_R \lbar r_R \lket$,
$\rbra r_L \rbar r_L \rket$ and  $\rbra \tilde p_\text{ref}=r_L \rbar r_L \rket$,
but also the Schouten identity, is used. As detailed
in \appref{sec:vertex_table}, this leads to many vanishing chirality-flow terms.
However, as opposed to the case of the (massless) fermion-gauge boson vertex
(cf. ref. \cite{Lifson:2022ijv} and \secref{sec:quark amps} below),
there is no speedup in the vertex subroutines themselves, which are
comparable in number of instructions to the native \mga\; implementation
(as measured using Valgrind \cite{Nethercote:2007vaf,Weidendorfer:2004ccs}).

\subsection{Processes with massless quarks}
\label{sec:quark amps}

\begin{figure}[t]
\begin{center}
  \includegraphics[scale=0.35]{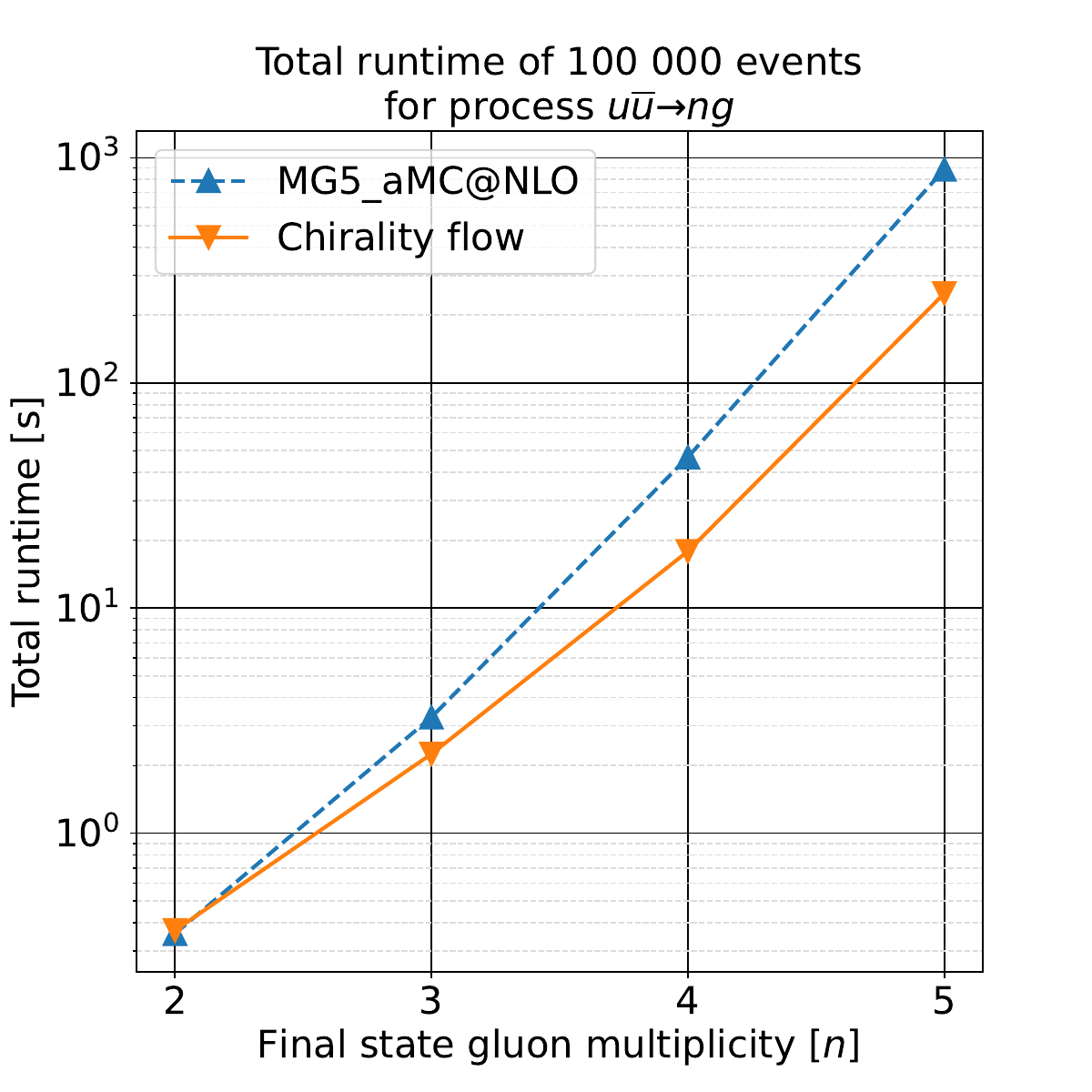}
  \includegraphics[scale=0.35]{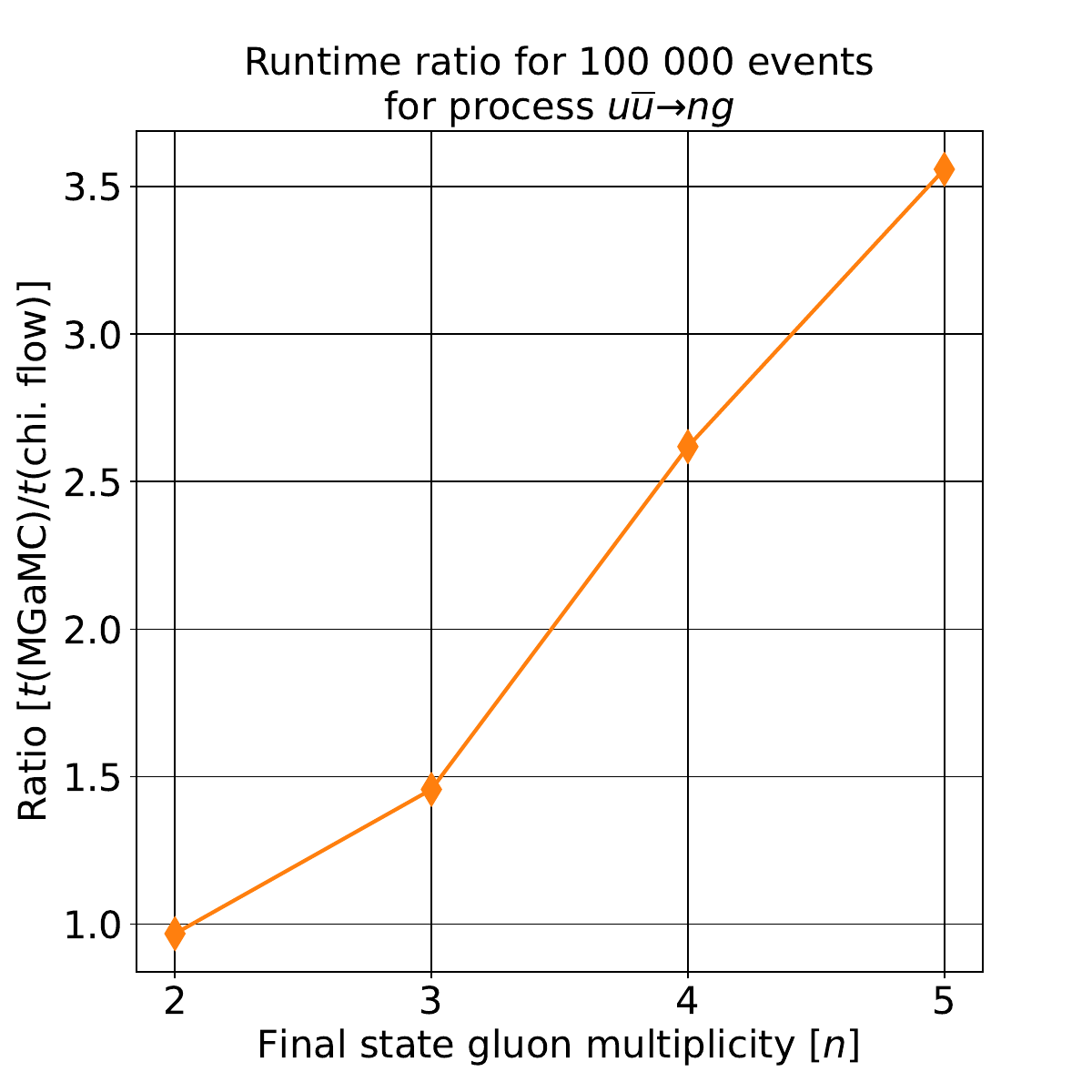}
  \caption{Runtime (left) and ratio (right) for the Lorentz structure for $u\ubar\rightarrow ng$ for the standalone version of \mga\; and  for the chirality-flow branch, cf.\,\figref{fig:gg_to_ng}. 
  } 
\label{fig:uu_to_ng}
\end{center}
\end{figure}

We next address amplitudes with one $\qqbar$ pair.
As discussed in \secref{sec:massless quarks}, we have found it
advantageous (for high multiplicities) to set reference
momenta according to the quarks, since this allows for the removal of
all diagrams where any left-chiral gluon is attached to the right-chiral
reference quark, and similarly for right-chiral gluons attached to the
left-chiral reference quark.

In \figref{fig:uu_to_ng} we display our results for $u\ubar \rightarrow ng$
(again with the squaring of color turned off, although here this effect is only
sizable for the five-gluon case).
As for the pure Yang-Mills case, we find an increased speedgain for high
multiplicities, but as opposed to the gluon case, this is not
only due to the gauge based (chirality-flow) diagram removal, but
also due to the compact nature of the fermion-gauge boson vertex,
as in ref. \cite{Lifson:2022ijv}.

\newpage
\subsection{Processes with massive quarks}
\label{sec:massive amps}

\begin{figure}[t]
  \begin{center}
    \includegraphics[scale=0.35]{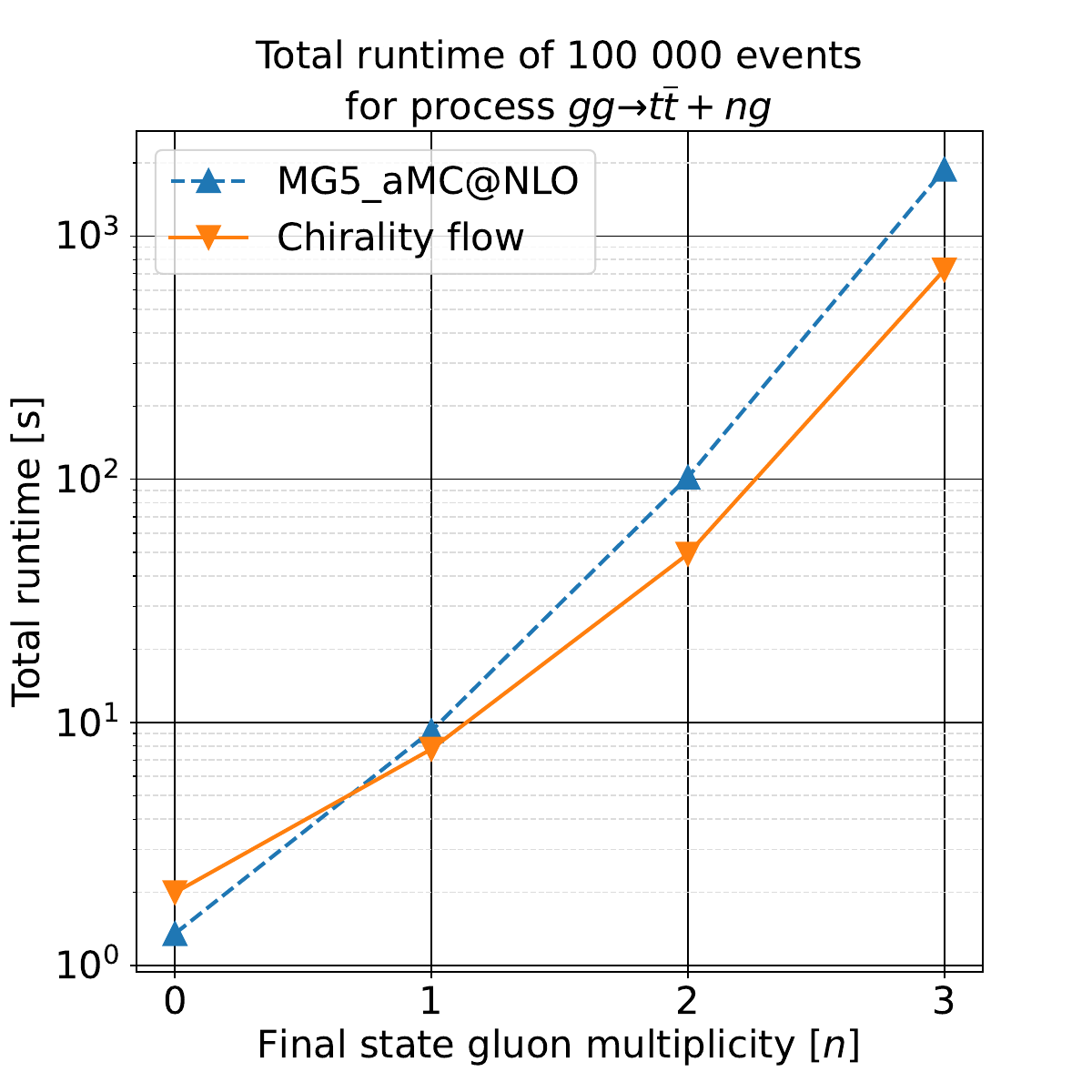}
    \includegraphics[scale=0.35]{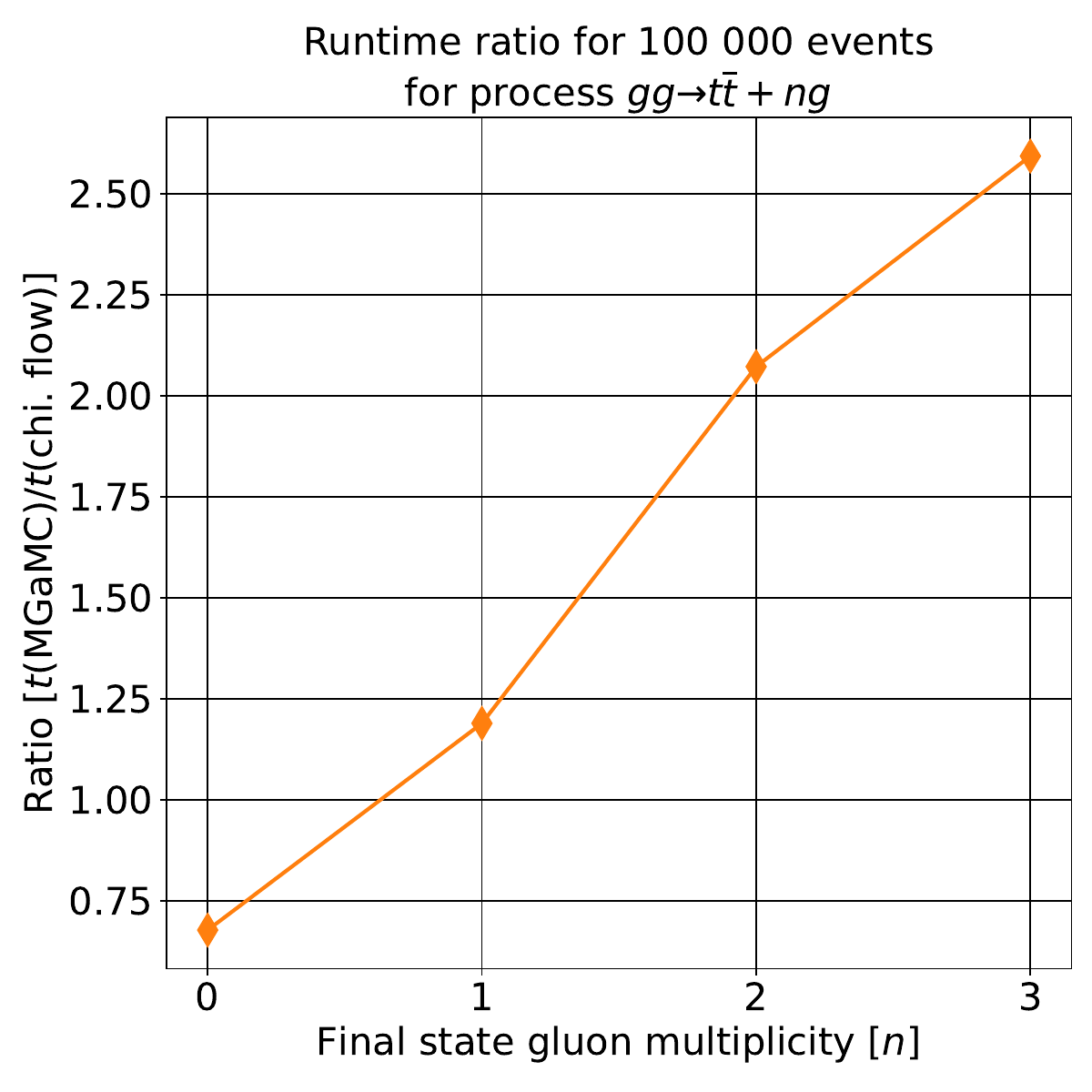}
    \caption{Runtime (left) and ratio (right) for the Lorentz structure of $gg \rightarrow t\bar t + ng$, for standalone \mga\;
      and for the chirality-flow version of it,  cf.\,\figref{fig:gg_to_ng}.
    	   }
    \label{fig:tt_to_ng}
  \end{center}
\end{figure}

As described in \secref{sec:QCD} we here have two kinds of vectors to
choose: The unphysical reference vectors of the gluons (which may in principle be picked
differently for processes with different fermion spins) and the physical
vector $q$, related to the spin direction as described in
\secref{sec:spin and gauge Feyn rules}.

The reference vector of a gluon can thus be set according to the
momentum of another gluon (as in the gluon-only case) or
according to one of the two spinors describing a massive quark.
In the latter case, we note that --- as opposed to for massless
quarks --- this will not remove the entire vertex (and thereby
Feynman diagram), but will only remove one chirality-flow
configuration. In the massive fermion case we have therefore 
found it
advantageous to set the reference momenta of the gluons
according to the momentum of a gluon with opposite chirality
(as for the pure Yang-Mills case).
The physical reference vector for the
quarks ($q$) is then 
set to equal the most frequently occurring gluon reference vector.

The exception to this is when all gluons have the same
chirality. In this case, the momentum $q$ for the
quarks is set to equal one gluon momentum, and the 
gluon reference momenta (all $r_L$ or $r_R$) are set to equal the
$p^\flat$ of the same quark.

A runtime comparison between the standalone version of \mga\;
and its chirality-flow branch is shown in
\figref{fig:tt_to_ng}.
Despite that this represents the worst possible case from the
chirality-flow perspective, as the fermions are not chiral,
and the vertices with gluons are naively not
faster than their native \mga\; versions, there is
an increasing speedgain. 
This is both due to vanishing chirality-flow contributions 
and the simplified massive fermion-gluon vertex. 

\subsection{Validation}

In order to ensure the accuracy of the chirality-flow implementation, validations were 
performed by comparing the squared amplitudes for several classes of processes to those 
obtained using default \mga. 
For each process, a randomly generated phase space point was evaluated
by both implementations for pure QCD. The spin summed squares of the 
amplitudes were found to be equal within numerical precision.
The processes included in the validation were:
\begin{itemize}
\item[(i)] Pure gluon: $gg\rightarrow ng$, $2\leq n \leq5$.
\item[(ii)] Massless quarks: $u\bar{u}\rightarrow ng$, $2\leq n \leq5$; 
$ug\rightarrow ug$; $ug\rightarrow ugg$.
\item[(iii)] Massive quarks: $gg\rightarrow t\bar{t}\,ng$, $0\leq n \leq3$; 
$t\bar{t}\rightarrow ng$, $2\leq n \leq5$; $tg\rightarrow tg$; $tg\rightarrow t\,gg$; 
$t\bar{t}\rightarrow t\bar{t}\,g$; $t\bar{t}\rightarrow t\bar{t}\,gg$; $gu\rightarrow t\bar{t}u$;
$u\bar{u}\rightarrow t\bar{t}\,g$; $u\bar{u}\rightarrow t\bar{t}\,gg$.
\end{itemize} 

\section{Conclusion and outlook}
\label{sec:conclusion}

The chirality-flow formalism represents a recast of the treatment
of the Lorentz structure giving reformulated chirality-flow
Feynman rules.
This parallels the construction of color-flow Feynman rules \cite{Kanaki:2000ms, Maltoni:2002mq},
with the difference that there are two types of flows, left-
and right-chiral, instead of just one color flow.
Furthermore, the chirality-flow lines necessarily
end in well-defined spinors for the external particles,
rather than unobserved color states which may be summed/averaged over.

This can be exploited to make many diagrams/terms vanish
by setting the gauge reference vectors for external gluons to
be the same for all left- and right-chiral gluons respectively,
and by further equating them to 
some other massless spinors in the
amplitude. In this way, simplified gauge-specific versions of
the gluon vertices can be defined, and many Feynman diagrams or contributions
to Feynman diagrams vanish as a spinor is contracted with itself.
Superficial simplifications from good choices of reference vectors 
are present in the spinor-helicity formalism itself,
but the flow representation
makes it obvious how to identify vanishing terms arbitrarily deep inside
Feynman diagrams.
The flow nature of chirality flow also lends itself to the
application of the Schouten identity already at the level
of the four-gluon vertex, 
allowing for further simplified versions
of it.

For massive quarks, we remark that the choice of reference vector carries
physical meaning, as it gives the direction in which
spin is measured. Thus, we here explore the freedom to measure
spin in arbitrary directions, rather than in the direction of motion
for each particle separately. Again, we find the flow description ideally
suited to shed light on this largely unexploited degree of freedom.

We have investigated the numerical speedup of scattering amplitude 
evaluations achieved in a 
\mga\; standalone implementation of chirality flow for QCD, including
massive quarks.
In short, we find an increasing speedgain for high multiplicities.
For the highest multiplicities, 
we find the Lorentz structure treatment to be
approximately a factor three faster, a
bit more for massless QCD, and a bit less for processes with a
$t\bar{t}$-pair.
Largely the acceleration
is due to the vanishing of either
entire Feynman diagrams or contributing chirality-flow diagrams.
These contributions are removed before compile time,
and tailored vertex subroutines, already encoding the simplifications,
are used at runtime. When applying helicity-recycling
\cite{Mattelaer:2021xdr} a similar setup can be used, with the
exception that the same reference vectors would have
to be used across processes of different helicity structure,
which would somewhat reduce the speedgain.

There is also some speedup due to the smaller
Lorentz structure of the quark-fermion vertex --- especially
in the massless case.
However,
as gluon vertices (which carry a similar number of instructions
as the ALOHA-generated versions)
tend to dominate for high multiplicities, this effect
is not as pronounced as in the massless QED implementation \cite{Lifson:2022ijv}.

In the present paper, we have addressed QCD, but it is worth
remarking that 
the weak interaction is in principle even better suited for
chirality flow, due to the chiral nature of the $W$-coupling.
Addressing the weak interaction is therefore a next natural step,
along with treating processes beyond tree level.

\section*{Acknowledgments}
We thank Olivier Mattelaer for productive discussions on 
the \mga{} implementation. We also thank Rikkert Frederix and
Olivier Mattelaer for constructive feedback on the manuscript. 
ZW thanks Stefan Roiser and Robert Sch\"ofbeck 
for enabling pursuit of this work. 
AL and MS acknowledge support by the Swedish Research Council (contract
number 2016-05996, as well as the European Union’s Horizon 2020
research and innovation programme (grant agreement No 668679).  The authors
have in part also been supported by the European Union’s Horizon
2020 research and innovation programme as part of the Marie
Sklodowska-Curie Innovative Training Network MCnetITN3 (grant
agreement no. 722104).  

\newpage

\appendix

\section{Additional chirality-flow rules}
\label{sec:additional rules}

The Dirac spinors for outgoing fermions $\bar{u}^\Js(p)$
and anti-fermions $v^\Js(p)$ in the chirality-flow picture may be
written as\,
\begin{align}
\bar{u}^+(p)
=
\raisebox{-5.5pt}{\includegraphics[scale=0.375]{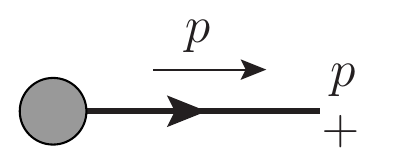}}
\hspace{-2ex}
&=
\bigg(
\!
\raisebox{-5.5pt}{\includegraphics[scale=0.375]{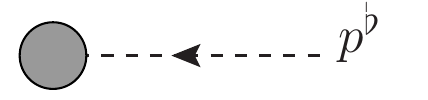}}
\hspace{-2ex}
\;,\;
-e^{i\varphi}
\sqrt{\al}
\raisebox{-5.5pt}{\includegraphics[scale=0.375]{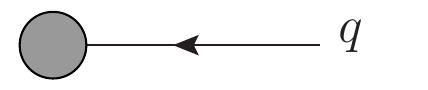}}
\hspace{-2ex}
\bigg)
\;, \nonumber
\\
\bar{u}^-(p)
=
\raisebox{-5.5pt}{\includegraphics[scale=0.375]{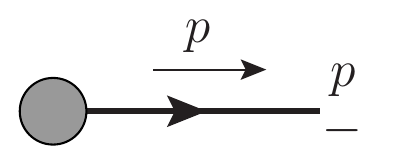}}
\hspace{-2ex}
&=
\bigg(
e^{\!-i\varphi}
\sqrt{\al}
\raisebox{-5.5pt}{\includegraphics[scale=0.375]{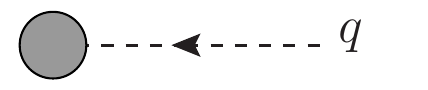}}
\hspace{-2ex}
\;,
\raisebox{-5.5pt}{\includegraphics[scale=0.375]{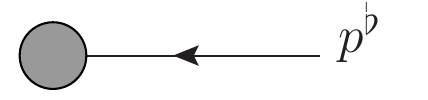}}
\hspace{-1.5ex}
\bigg)
\;, \nonumber
\\
v^+(p)
=
\raisebox{-5.5pt}{\includegraphics[scale=0.375]{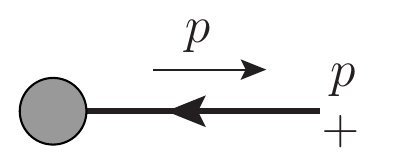}}
\hspace{-2ex}
&=
\begin{pmatrix}
\phantom{\,-e^{i\varphi}\!\sqrt{\alpha}}
\raisebox{-5.5pt}{\includegraphics[scale=0.375]{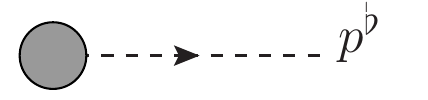}}
\hspace{-1.25ex}
\\
\,-e^{i\varphi}\!\sqrt{\alpha}
\raisebox{-5.5pt}{\includegraphics[scale=0.375]{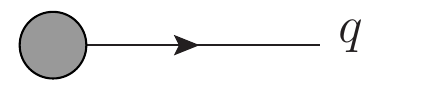}}
\hspace{-1.25ex}
\end{pmatrix}
\;, \nonumber
\\
v^-(p)
=
\raisebox{-5.5pt}{\includegraphics[scale=0.375]{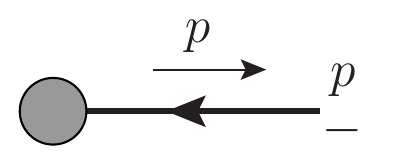}}
\hspace{-2ex}
&=
\begin{pmatrix}
e^{-i\varphi}\!\sqrt{\alpha}
\raisebox{-5.5pt}{\includegraphics[scale=0.375]{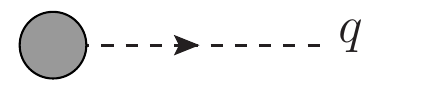}}
\hspace{-1.75ex}
\\
\phantom{e^{-i\varphi}\!\sqrt{\alpha}}
\raisebox{-5.5pt}{\includegraphics[scale=0.375]{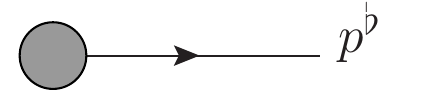}}
\hspace{-1.75ex}
\end{pmatrix}
\;,
\label{eq:ubarvpm_spin_cf}
\end{align}
where in the massless limit $p^\flat=p$ and $\alpha = 0$.
The factors 
\begin{equation}
  e^{i\varphi}\sqrt{\alpha}=\frac{m}{\cAngle{p^\flat q}}\;,
  \qquad 
  e^{\!-i\varphi}\sqrt{\alpha}=\frac{m}{\cSquare{q p^\flat}}~, \label{eq:spinor phase def with alpha}
\end{equation}
are required such that the spinors satisfy the Dirac equation, and
the phase is given by
\begin{equation}
  \cSquare{q p^\flat} =e^{i\varphi}\sqrt{2p^\flat\!\cdot q}\;,
  \qquad 
  \cAngle{p^\flat q} = e^{-i\varphi}\sqrt{2p^\flat\!\cdot q}~. \label{eq:spinor phase def}
\end{equation}
Explicit forms of the Weyl spinors (fixing the phase $\varphi$),
as well as other conventions can be found in ref. \cite{Alnefjord:2020xqr}.

Similarly, for incoming anti-fermions $\bar{v}^\Js(p)$ and fermions $u^\Js(p)$ we write
\begin{align}
\bar{v}^+(p)
&=
\raisebox{-5.5pt}{\includegraphics[scale=0.375]{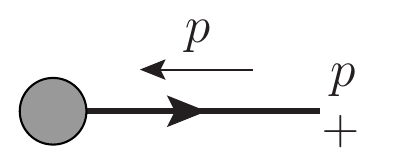}}
\hspace{-2ex}
=
\bigg(
-e^{\!-i\varphi}\!\sqrt{\alpha}
\raisebox{-5.5pt}{\includegraphics[scale=0.375]{./Jaxodraw/CR/ExtSpinorDotted_q}}
\hspace{-2ex}
\;,\;
\raisebox{-5.5pt}{\includegraphics[scale=0.375]{./Jaxodraw/CR/ExtSpinorSolid_pflat}}
\hspace{-1.5ex}
\bigg)
\;, \nonumber
\\
\bar{v}^-(p)
&=
\raisebox{-5.5pt}{\includegraphics[scale=0.375]{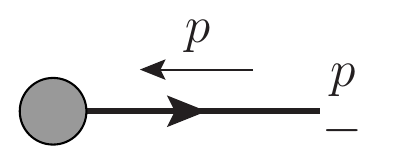}}
\hspace{-2ex}
=
\bigg(
\!\raisebox{-5.5pt}{\includegraphics[scale=0.375]{./Jaxodraw/CR/ExtSpinorDotted_pflat}}
\hspace{-2ex}
\;,\;
e^{i\varphi}\!\sqrt{\alpha}
\raisebox{-5.5pt}{\includegraphics[scale=0.375]{./Jaxodraw/CR/ExtSpinorSolid_q}}
\hspace{-2ex}
\bigg)
\;, \nonumber
\\
u^+(p)
&=
\raisebox{-5.5pt}{\includegraphics[scale=0.375]{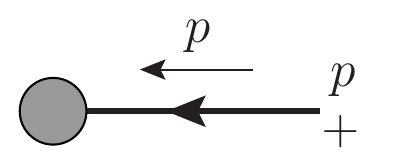}}
\hspace{-2ex}
=
\begin{pmatrix}
-e^{\!-i\varphi}\!\sqrt{\alpha}
\raisebox{-5.5pt}{\includegraphics[scale=0.375]{./Jaxodraw/CR/ExtSpinorAntiDotted_q}}
\hspace{-1.75ex}
\\
\phantom{-e^{\!-i\varphi}\!\sqrt{\alpha}}
\raisebox{-5.5pt}{\includegraphics[scale=0.375]{./Jaxodraw/CR/ExtSpinorAntiSolid_pflat}}
\hspace{-1.75ex}
\end{pmatrix}
\;, \nonumber
\\
u^-(p)
&=
\raisebox{-5.5pt}{\includegraphics[scale=0.375]{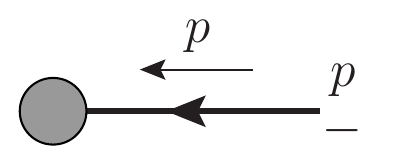}}
\hspace{-2ex}
=
\begin{pmatrix}
\phantom{e^{i\varphi}\!\sqrt{\alpha}}
\raisebox{-5.5pt}{\includegraphics[scale=0.375]{./Jaxodraw/CR/ExtSpinorAntiDotted_pflat}}
\hspace{-1.25ex}
\\
e^{i\varphi}\!\sqrt{\alpha}
\raisebox{-5.5pt}{\includegraphics[scale=0.375]{./Jaxodraw/CR/ExtSpinorAntiSolid_q}}
\hspace{-1.25ex}
\end{pmatrix}
\;.
\label{eq:uvbarpm_spin_cf}
\end{align}
In \eqsrefa{eq:ubarvpm_spin_cf}{eq:uvbarpm_spin_cf} 
the leftmost graphical rules correspond to the conventional Feynman rules 
(showing the fermion-flow arrows, momentum labels and spin labels; the thicker lines imply massive particles), 
and the rightmost graphical rules correspond to the chirality-flow rules.
Explicit spinors can be found in appendix B.1 in ref. \cite{Alnefjord:2020xqr}.

\section{Simplifications in three- and four-gluon vertices}
\label{sec:vertex_table}

\begin{table}[t]
    \centering
    \begin{tabular}{cc|cc} \hlinewd{1pt}
       3-gluon vertex  & No. structures & 3-gluon vertex & No. structures  \\ \hline
       $g_{LR} \, g_{LR} \, g_{LR}$ & 0 & $g_R \, g_R \, g_R$ & 0 \\ 
       $g_{LR} \, g_{LR} \, g_R$ & 0 & $g_R \, g_R \, g_L$ & 2 \\ 
       $g_{LR} \, g_{LR} \, g$ & 2 & $g_R \, g_R \, g$ & 2 \\ 
       $g_{LR} \, g_R \, g_R$ & 0 & $g_R \, g_L \, g$ & 3 \\ 
       $g_{LR} \, g_R \, g_L$ & 1 & $g_R \, g \, g$ & 3 \\ 
       $g_{LR} \, g_R \, g$ & 2 & $g \, g \, g$ & 3 \\ 
       $g_{LR} \, g \, g$ & 3 & & \\ \hlinewd{1pt}
       4-gluon vertex  & No. structures & 4-gluon vertex & No. structures  \\ \hline
       $g_{LR} \, g_{LR} \, g_{LR} \, g_{LR} $ & 0 & $g_R \, g_R \, g_R \, g_R$ & 0 \\
       $g_{LR} \, g_{LR} \, g_{LR} \, g_R$ & 0 & $g_R \, g_R \, g_R \, g_L$ & 0 \\ 
       $g_{LR} \, g_{LR} \, g_{LR} \, g$ & 0 & $g_R \, g_R \, g_R \, g$ & 0 \\ 
       $g_{LR} \, g_{LR} \, g_R \, g_R$ & 0 & $g_R \, g_R \, g_L \, g_L$ & 1 \\ 
       $g_{LR} \, g_{LR} \, g_R \, g_L$ & 0 & $g_R \, g_R \, g_L \, g$ & 2 \\
       $g_{LR} \, g_{LR} \, g_R \, g$ & 0 & $g_R \, g_R \, g \, g$ & 2 \\
       $g_{LR} \, g_{LR} \, g \, g$ & 2 & $g_R \, g_L \, g \, g$ & 3 \\ 
       $g_{LR} \, g_R \, g_R \, g_R$ & 0 & $g_R \, g \, g \, g$ & 3 \\ 
       $g_{LR} \, g_R \, g_R \, g_L$ & 0 & $g \, g \, g \, g$ & 3 \\ 
       $g_{LR} \, g_R \, g_R \, g$ & 0 & & \\ 
       $g_{LR} \, g_R \, g_L \, g$ & 1 & & \\
       $g_{LR} \, g_R \, g \, g$ & 2 & & \\ 
       $g_{LR} \, g \, g \, g$ & 3 & & \\ \hlinewd{1pt}
    \end{tabular}
    \caption{Gauge specific Feynman rules for the gluon vertices,
      with the first and third columns detailing the (explicitly chiral)
      particle configurations, and the second and fourth columns enumerating
      non-vanishing Lorentz structures for the corresponding configuration.
      Missing vertex configurations are obtained by applying a chirality flip,
      $L \leftrightarrow R$.
    }
    \label{tab:table_of_amp_rules}
\end{table}

\begin{table}[t]
    \centering
    \begin{tabular}{ccc|ccc} \hlinewd{1pt}
       Incoming &  Propagator & No. structures & Incoming & Propagator & No. structures \\ \hline
        $g_{LR} \, g_{LR}$ & $g_{LR}$ & 2 & $g_R \, g_R$ & $g_R$ & 2 \\ 
        $g_{LR} \, g_R$ & $g_R$ & 2 & $g_R \, g_L$ & $g$ & 3 \\ 
        $g_{LR} \, g$ & $g$ & 3 & $g_R \, g$ & $g$ & 3 \\ 
                & & & $g \, g$ & $g$ & 3 \\ \hlinewd{1pt} 
       Incoming &  Propagator & No. structures & Incoming & Propagator & No. structures \\ \hline
        $g_{LR} \, g_{LR} \, g_{LR}$ & 0 & 0 & $g_R \, g_R \, g_R$ & 0 & 0 \\ 
       $g_{LR} \, g_{LR} \, g_R$ & 0 & 0 & $g_R \, g_R \, g_L$ & $g_R$ & 2 \\ 
       $g_{LR} \, g_{LR} \, g$ & $g_{LR}$ &  1 & $g_R \, g_R \, g$ & $g_R$ & 2 \\
       $g_{LR} \, g_R \, g_R$ & 0 & 0 & $g_R \, g_L \, g$ & $g$ & 3 \\
       $g_{LR} \, g_R \, g_L$ & $g_{LR}$ & 1 & $g_R \, g \, g$ & $g$ & 3 \\ 
       $g_{LR} \, g_R \, g$ & $g_R$ & 2 & $g \, g \, g$ & $g$ & 3 \\ 
       $g_{LR} \, g \, g$ & $g$ & 3 & & & \\ \hlinewd{1pt}
    \end{tabular}
    \caption{Gauge specific Feynman for use in the HELAS algorithm
      described in \secref{sec:fhelas}, with the first and fourth columns detailing
      the (explicitly chiral) particle configurations, the second and fifth columns
      showing the resulting propagator, and the third and sixth columns enumerating
      non-vanishing Lorentz structures for the corresponding configuration.
      Missing vertex configurations are obtained by $L\leftrightarrow R$.
      Note that vertices with explicitly chiral particles as arguments return
      explicitly chiral propagators, with the sole exception of the
      $g_R \, g_L \, g$-vertex.}
    \label{tab:table_of_prop_rules}
\end{table}

In this appendix we list the simplifications in the three- and four-gluon
vertices. We first, in \tabref{tab:table_of_amp_rules}, address the gauge
dependent Feynman rules in the case when all gluons have a known chirality
status ($g_{LR}$, $g_L$, $g_R$ or $g$). In the context of the HELAS
algorithm, this is applicable to the last step of the evaluation.
We then, in \tabref{tab:table_of_prop_rules}, address the case when two or three gluons give rise to an off-shell
wavefunction, used in the first steps in the Feynman diagram evaluation.

\private{
\newpage
\section{Remaining TODOs and collected comments}

\subsection{TODOs}

  \todo{Emil validation}

  \todo{EB(?): Check eqs modified by MS}

  \todo{IN THE VERY END}
  \todo{Borows through}
  \todo{Final spell check}
  \todo{MCnet pre-print number}
  
  \subsection{Other comments}

  ----------------------------DON'T FORGET-----------------------------

\ZW{note: not yet implemented diagram removal at export step. idea going forward: 
remove non-contributing vertices either in helas\_objects.py (HelasWavefunction, HelasAmplitude, (HelasDiagram?)) 
or in helas\_call\_writers.py (string manipulation based on names of vertex function calls }

\todo{doubly chiral gluons: same kinematic amplitude sometimes multiplies several color structures}

\ZW{metric propagator and reuse of amplitudes (for future implementations): 
consider rewriting code so that reused values are called by reference instead 
of copied in memory? probably no major gain, but I want to mention it so 
we all think about it}

irrespective of if the vertex results in an off-shell wavefunction or an amplitude.

{\red \todo{From the example. Note that it can output a chiral fermion. Show with an example how it helps the rest of the diagram, i.e.\ it makes some later vertex disappear, e.g.\ the fermion-vector vertex. Ex gLgLg-vertex with g attaching to left fermion }}
}

\newpage
\bibliographystyle{JHEP}  
\bibliography{refs} 

\end{document}